\documentclass[preprint,11pt]{aastex}

\newcommand{\FAC}{HE~1327$-$2326} 
\newcommand{\CBB}{HE~0107$-$5240} 
\newcommand{\GG}{G~64$-$12} 
\newcommand{\cd}{CD~$-38^{\circ}\,245$}
\newcommand{\tefft}{$T_{\mbox{\scriptsize eff}}$}
\newcommand{\teffm}{T_{\mbox{\scriptsize eff}}}
\newcommand{\vmicrt}{$v_{\mbox{\scriptsize micr}}$}
\newcommand{\vmicrm}{v_{\mbox{\scriptsize micr}}}

\shorttitle{Abundance analysis of {\FAC}}
\shortauthors{Aoki et al.}

\begin{document}

\title{{\FAC}, an unevolved star with $\mbox{[Fe/H]}<-5.0$. \\
       I. A Comprehensive Abundance Analysis\altaffilmark{1}}

%
%

\author{
  W. Aoki\altaffilmark{2,3,4},
  A. Frebel\altaffilmark{2,5},
  N. Christlieb\altaffilmark{6,3},
  J.E. Norris\altaffilmark{5},
  T.C. Beers\altaffilmark{7},
  T. Minezaki\altaffilmark{8,9}
  P.S. Barklem\altaffilmark{10},
  S. Honda\altaffilmark{3},
  M. Takada-Hidai\altaffilmark{11},
  M. Asplund\altaffilmark{5},
  S.G. Ryan\altaffilmark{12},
  S. Tsangarides\altaffilmark{12},
  K. Eriksson\altaffilmark{10},
  A. Steinhauer\altaffilmark{13,14},
  C. P. Deliyannis\altaffilmark{15},
  K. Nomoto\altaffilmark{16},
  M.Y. Fujimoto\altaffilmark{17},
  H. Ando\altaffilmark{3},
  Y. Yoshii\altaffilmark{8,9},
  T. Kajino\altaffilmark{3,4},
}

\altaffiltext{1}{Based on data collected with the Subaru Telescope, which is
  operated by the National Astronomical Observatory of Japan.}
\altaffiltext{2}{The first two authors have contributed equally to the results
  presented in this paper.}
\altaffiltext{3}{National Astronomical Observatory of Japan, 2-1-21
  Osawa, Mitaka, Tokyo, 181-8588 Japan; aoki.wako@nao.ac.jp,
  honda@optik.mtk.nao.ac.jp, ando@optik.mtk.nao.ac.jp,
  kajino@nao.ac.jp}
\altaffiltext{4}{Department of Astronomy, Graduate University of Advanced
  Studies, Mitaka, Tokyo 181-8588, Japan}
\altaffiltext{5}{Research School of Astronomy and Astrophysics, Australian 
National University, Cotter Road, Weston, ACT 2611, Australia;
    anna@mso.anu.edu.au, jen@mso.anu.edu.au, martin@mso.anu.edu.au}
\altaffiltext{6}{Hamburger Sternwarte, University of Hamburg, Gojenbergsweg
  112, D-21029 Hamburg, Germany; nchristlieb@hs.uni-hamburg.de}
\altaffiltext{7}{Department of Physics and Astronomy and JINA: Joint
  Institute for Nuclear Astrophysics, Michigan State University, East
  Lansing, MI 48824, USA; beers@pa.msu.edu}
\altaffiltext{8}{Institute of Astronomy, School of Science, University
  of Tokyo, Mitaka, Tokyo 181-0015, Japan;
  minezaki@kiso.ioa.s.u-tokyo.ac.jp, yoshii@ioa.s.u-tokyo.ac.jp}
\altaffiltext{9}{Research Center for the Early Universe, School of Science,
  University of Tokyo, Bunkyo-ku, Tokyo 113-0033, Japan}
\altaffiltext{10}{Uppsala Astronomical Observatory, Box 515, SE-75120 Uppsala, 
  Sweden; barklem@astro.uu.se, Kjell.Eriksson@astro.uu.se,}
\altaffiltext{11}{Liberal Arts Education Center, Tokai University,
  1117 Kitakaname, Hiratsuka-shi, Kanagawa 259-1292, Japan;
  hidai@apus.rh.u-tokai.ac.jp}
\altaffiltext{12}{Department of Physics and Astronomy, Open
University, Walton Hall, Milton Keynes MK76AA, UK;
S.G.Ryan@open.ac.uk, S.Tsangarides@open.ac.uk}
\altaffiltext{13}{Department of Astronomy, University of Florida, 211
Bryant Space Science Center, Gainesville, FL 32611-2055}
\altaffiltext{14}{Present address: SUNY Geneseo, Department of Physics
and Astronomy, One College Circle, Geneseo, NY 14454,
steinhau@geneseo.edu}
\altaffiltext{15}{Department of Astronomy, Indiana University, 727
East 3rd Street, Swain Hall West 319, Bloomington, IN 47405-7105
con@athena.astro.indiana.edu}
\altaffiltext{16}{Department of Astronomy, School of Science, University of
  Tokyo, Tokyo 113-0033, Japan: nomoto@astron.s.u-tokyo.ac.jp}
\altaffiltext{17}{Department of Physics, Hokkaido University, Sapporo
  060-0810, Japan; fujimoto@astro1.sci.hokudai.ac.jp}

\begin{abstract} 

We present the elemental abundances of {\FAC}, the most iron-deficient
star known, determined from a comprehensive analysis
of spectra obtained with the Subaru Telescope High Dispersion
Spectrograph. {\FAC} is either in its main sequence or subgiant phase
of evolution. Its NLTE corrected iron abundance is [Fe/H]$=-5.45$,
0.2~dex lower than that of {\CBB}, the previously most iron-poor
object known, and more than 1~dex lower than those
of all other metal-poor stars. Both {\FAC} and {\CBB} exhibit
extremely large overabundances of carbon ([C/Fe]$\sim +4$). The
combination of extremely high carbon abundance with outstandingly low
iron abundance in these objects clearly distinguishes them from other
metal-poor stars. The large carbon excesses in these two stars are not
the result of a selection effect.

There also exist important differences between {\FAC} and
{\CBB}. While the former shows remarkable overabundances of the light
elements (N, Na, Mg and Al), the latter shows only relatively small
excesses of N and Na. The neutron-capture element Sr is detected in
{\FAC}, but not in {\CBB}; its Sr abundance is significantly higher
than the upper limit for {\CBB}. The \ion{Li}{1} 6707~{\AA} line,
which is detected in most metal-poor dwarfs and warm subgiants having
the same temperature as {\FAC}, is not found in this object. The upper
limit of its Li abundance ($\log\epsilon\left(\mbox{Li}\right) < 1.5$)
is clearly lower than the Spite plateau value.

These data provide new constraints on models of nucleosynthesis
processes in the first generation objects that were responsible for
metal enrichment at the earliest times. We discuss possible scenarios
to explain the observed abundance patterns.

\end{abstract}


\section{INTRODUCTION}

Recent simulations to study the formation of the first generations of
stars predict the formation of objects with masses from several
hundred solar mass (M$_{\odot}$) to as large as
1000~M$_{\odot}$. These massive stars are believed to have played a
crucial role in the re-ionization of the early Universe and metal
enrichment \citep[][and references therein]{bromm04}.  In particular,
supermassive stars ($>130$ M$_{\odot}$), which terminate their lives
as pair-instability supernovae or direct formation of black holes,
would be a unique type of objects formed only from primordial clouds
\citep[e.g. ][]{Heger/Woosley:2002}. In contrast, the formation of
low-mass stars from metal-free clouds has been a long-standing
problem. While some studies suggest that their formation might have
been prohibited due to the lack of an efficient cooling source such as
dust grains during the collapse of gas clouds, others predict the
production of objects having masses $\sim 1$M$_{\odot}$ \citep[][ and
references therein]{bromm04}. Although the mass function of these
first objects is a key to estimating their contribution to
re-ionization, metal enrichment, and subsequent formation of stars and
galaxies, no definitive observational constraint has yet been made.

Elemental abundance measurements for the most metal-poor stars found
in our Galaxy have played a unique role in studies of the first
stellar generations, because the record of their nucleosynthetic
yields is believed to be preserved in the atmospheres of extremely
metal-poor stars.  The large objective-prism surveys for metal-poor
stars that have been conducted over the past two decades, the HK
survey of Beers and colleagues (Beers et al. 1985, 1992; Beers 1999),
and the Hamburg/ESO survey of Christlieb and colleagues
\citep{Christlieb:2003}, have provided substantial samples of very
metal-poor stars, but no object with [Fe/H]$<-4$ was identified until
quite recently\footnote{[A/B] = $\log(N_{\rm A}/N_{\rm B})-
\log(N_{\rm A}/N_{\rm B})_{\odot}$, and $\log \epsilon_{\rm A} =
\log(N_{\rm A}/N_{\rm H})+12$ for elements A and B.}.  The discovery
of {\CBB}, a giant with $\mbox{[Fe/H]}=-5.3$ \citep{HE0107_Nature,
HE0107_ApJ} was a breakthrough in the study of the first generation of
stars and supernovae (SNe), and the star formation processes that were
operating in the early Universe. In contrast to its low Fe abundance,
{\CBB} exhibits extreme overabundances of carbon, nitrogen, and oxygen
(Christlieb et al. 2002; Bessell et al. 2004).

A number of possibilities have been proposed to account for the
distinctive abundance pattern of elements in {\CBB}. Several models
assume that it is a second generation object, and explain its
abundance pattern by calling for pollution by either one or two
supernovae
\citep{Umeda/Nomoto:2003,Limongietal:2003}. \citet{Shigeyamaetal:2003}
investigated possible accretion of material from the interstellar
matter (ISM), and suggested that this star may be a low-mass first
generation object. \citet{suda04} investigated this possibility in
more detail, combining the production of light elements (e.g. C, N)
in an AGB star with mass transfer across a binary system.  More
detailed discussions of these models are presented in
\citet{beers05}. Clearly, there is no present consensus on the origin
of the abundance pattern of {\CBB}, and further observational studies
are needed to place stronger constraints on the various proposed
scenarios.

As has been reported by \citet{frebel05},
spectra of the subgiant or dwarf {\FAC} obtained with the Subaru
Telescope High Dispersion Spectrograph \citep[HDS: ][]{noguchi02} in
May and June 2004 revealed that its Fe abundance is even lower than
that of {\CBB}. Aside from its obvious importance in establishing the
class of Hyper Metal-Poor \citep[HMP: ][]{beers05} stars with [Fe/H]
$< -5.0$, the chemical abundance pattern of {\FAC} provides additional
important constraints on the models proposed for {\CBB}. In \citet{frebel05},
we presented results of an abundance analysis based on the Subaru/HDS
spectra.  Here we describe our observations and analysis in more
detail, and discuss more extensively the implications of the derived
abundances, as well as possible scenarios for the origin of the
abundance pattern of {\FAC}. In \S 2 we report details of the sample
of stars from which {\FAC} was identified, as well as the observations
and measurements obtained to date. The determination of atmospheric
parameters is described in \S 3.  In \S 4 we present the abundance and
error analysis. An overview of the derived elemental abundances and a
summary of the abundance characteristics for this star is presented in
\S 5. Possible interpretations of the observed abundances are
discussed in \S 6, while the implications for future surveys to
identify additional HMP stars are given in \S 7. A summary and
concluding remarks are presented in \S 8.


\section{OBSERVATIONS AND MEASUREMENTS}

\subsection{Sample selection}

The digital spectra of the Hamburg/ESO objective-prism survey
\citep[HES;][]{hespaperIII} has yielded numerous candidate faint
metal-poor stars over the past years. {\CBB}, a $B=15.86$\,mag star,
was found amongst these faint objects \citep{HE0107_ApJ}.

The selection of candidates for extremely metal-poor stars
\citep{Christlieb:2003} was recently extended to the brightest objects
that are measured in the HES; that is, to the magnitude range
$10\lesssim B\lesssim 14$ (Frebel et al., in preparation).  The HES
objective-prism spectra of stars brighter than $B\sim 14$ suffer from
saturation effects, and they were thus excluded from the original
selection procedure.  However, careful investigation has shown that
viable metal-poor candidates can still be selected from among these
bright objects, albeit with a 3--4 times lower efficiency than for the
selection of fainter stars. {\FAC} is one of the 1777 bright
metal-poor candidates found on 329 (out of a total of 380) HES plates,
covering a nominal area of 8225 square degrees of the southern
high-galactic latitude sky.

In the course of moderate-resolution ($\sim 2$\,{\AA}) follow-up
observations of the entire sample, {\FAC} stood out because of its
very weak \ion{Ca}{2} K line.  In a medium-resolution spectrum
obtained with the ESO 3.6\,m telescope and EFOSC, the \ion{Ca}{2} K
line index \texttt{KP} and the H$\delta$ index \texttt{HP2}
\citep{Beersetal:1999} were measured to be $0.26$ {\AA} and $3.66$
{\AA}, respectively, yielding $\mbox{[Fe/H]}=-4.0$, when the
calibration of these indices of \cite{Beersetal:1999} was employed. A
second, longer exposure medium-resolution spectrum was taken in May
2004 with the Double Beam Spectrograph at the SSO 2.3~m
telescope. Measurements of \texttt{KP} $=0.16$ {\AA} and \texttt{HP2}
$=4.01$ {\AA} refined the previously derived metallicity to
$\mbox{[Fe/H]}=-4.3$. The $B-V$ color and HP2 value of {\FAC} appeared
to be similar to that of CS~22876--032, which was the most metal-poor
dwarf known at that time ($\mbox{[Fe/H]}=-3.71$: Norris et al. 2000). 
This suggested that the effective temperatures of the two stars were
very similar; further comparison indicated that the metallicity of
{\FAC} might indeed be lower than that of CS~22876--032.

We have been conducting an observing program to measure chemical abundance
patterns of ultra metal-poor stars with Subaru/HDS (P.I.: Aoki) since semester
S03B (2003). {\FAC} became a high-priority target during 2004. For comparison
purposes, high-resolution data were also obtained for the well-studied,
extremely metal-poor dwarf {\GG}\footnote{We note that this star was re-identified among
the bright HES candidates as very metal-poor, and had been designated
HE~1337+0012.}, which has temperature and gravity similar
to those of {\FAC}.

\subsection{High-resolution spectroscopy}

High-resolution spectra of {\FAC} were obtained with Subaru/HDS during
four nights in May and June 2004 using three different settings,
covering a total wavelength range of 3050--6800\,{\AA}. Comparison
spectra of {\GG} were obtained with two settings, covering
3050--5250~{\AA}. Details of the observations are listed in Table
\ref{Tab:SubaruObs}. The slit width was 0.6~arcsec, yielding a
resolving power of $R=60,000$. CCD on-chip binning (2 x 2 pixels) was
applied.

After bias subtraction and multiplication by the gain factor, the
spectra were reduced independently by the first two authors in
slightly different ways. W.A. used IRAF\footnote{IRAF is distributed
by the National Optical Astronomy Observatories, which is operated by
the Association of Universities for Research in Astronomy, Inc. under
cooperative agreement with the National Science Foundation.}
procedures, while A.F.  used Figaro \citep{figaro}. 
The main difference was that W.A. co-added the individual exposures
for a given setting by summing extracted, wavelength-calibrated, and
Doppler-corrected spectra. Cosmic-ray hits in the two dimensional CCD
images were removed as described in \citet{aoki05}. A.F., however,
computed the median of the frames to account for cosmic ray removal
before extracting the orders.

After order merging, it was noticed that the final spectrum reduced by
W.A. exceeded the quality of the spectrum reduced by A.F., presumably
due to the different methods of co-adding the
individual observations. The $S/N$ of W.A.'s spectrum is 20/1 at
3150\,{\AA}, 160/1 at 4000\,{\AA} and 170/1 at 6700\,{\AA}. The $S/N$ is
highest (260/1) at 4600~{\AA}, because this wavelength was covered by
all three setups. The corresponding values for the spectrum of {\GG}
are 50/1 at 3150~{\AA}, 220/1 at 4000~{\AA}, and 270/1 (highest) at
4600~{\AA}.  Hence, we decided to mainly work with the better, higher $S/N$
spectrum but to keep using the spectrum reduced by A.F. for
consistency checks on our measurements, as originally intended. In the
following, telluric absorption at redder wavelengths was corrected
only for the spectrum with higher $S/N$ by using the spectrum of a
standard star, HD~114376 (B7 III), obtained with the same setup
employed for {\FAC}.  We note that the superiority of one spectrum is
the reason why only the strongest lines could be measured in the
spectrum by A.F., and that no upper limits for elements were intended
to be derived from that spectrum.

\begin{figure*}[phtb]
  \includegraphics[width=9cm]{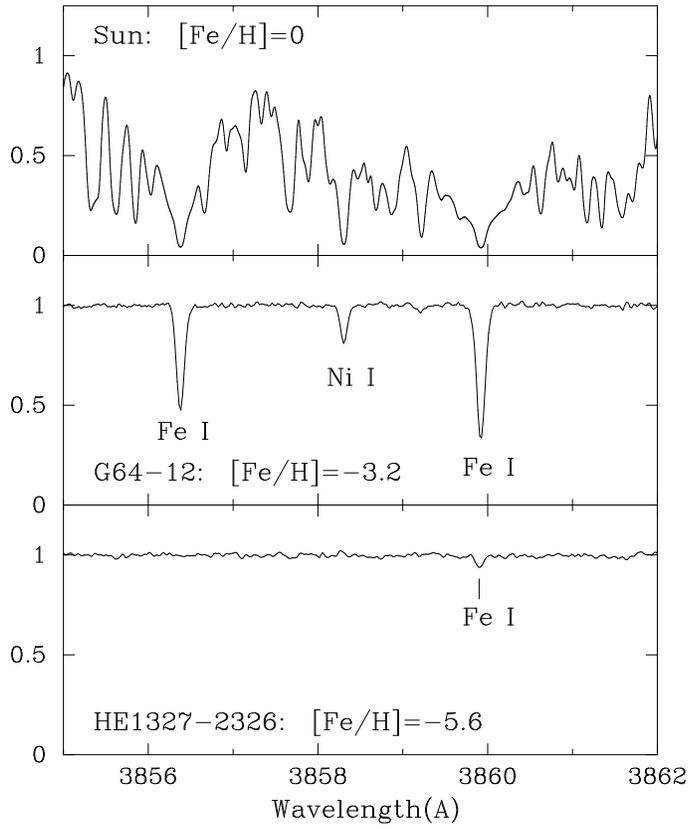}
  \caption{\label{Fig:FeLines} The \ion{Fe}{1} 3860~{\AA} line in the
  spectrum of {\FAC} (bottom), compared with spectra of the Sun (top)
  and {\GG} (middle).}

\end{figure*}

Examples of several spectral regions of the higher $S/N$ spectrum are shown in
Fig.~\ref{Fig:FeLines}, together with those of {\GG} and the
Sun. This figure includes the \ion{Fe}{1} line at 3860~{\AA} detected
in {\FAC}, which is much weaker than in {\GG}.

\subsection{Line identification and measurements of equivalent
widths}\label{sec:ew}

In the spectrum of {\FAC} only a very small number of atomic lines is
seen, while a number of weak CH and NH molecular lines appear, as
reported in \citet{frebel05}. We systematically searched for atomic
lines known to be relatively strong in other metal-poor stars,
referring to previous studies in the literature. For iron-peak
elements (Ti to Zn), for which a large number of absorption lines are
usually detected even in extremely metal-poor stars, we listed the
strongest lines measured for {\GG}, and carefully looked for them in
the spectrum of {\FAC}. As a result, we found a total of 23 lines,
among which 17 are used to determine the final abundances (see
\S~\ref{sec:ana}).

Equivalent widths of the absorption lines of {\FAC} were measured by
fitting Gaussian profiles. Where possible, measurements were made
independently by W.A. and A.F. in their reduced spectra. The two sets
of values were found to agree within the measurement uncertainties
(see Table \ref{Tab:Eqw}). For each line, we adopt the average of the
two measurements when the lines are detected in both sets of the
spectra.  Otherwise, the equivalent widths measured by W.A are
used. We note that the \ion{Na}{1} lines, which are relatively strong
in this object, were measured only in the spectrum reduced by W.A.,
who applied a correction for the contamination of telluric
absorption. The results are listed in Table \ref{Tab:Eqw}.


Upper limits on the equivalent widths for several elements that are
not detected in our spectrum were estimated using the formula of
\citet{Norrisetal:2001} for the S/N ratio at the wavelength of each
line. The upper limits on the equivalent widths given in Table
\ref{Tab:Eqw} are 3$\sigma$ values. For undetected species we give data
for the line that provides the lowest upper limit on the abundance.

\subsection{Line widths}\label{sec:linewidth}

The widths of several lines were investigated in order to check if {\FAC} has
anomalously high rotational velocity or macroturbulence, which is related to the
discussion of the Li abundance (\S~\ref{sec:lidep}). In the spectrum of {\FAC},
only two \ion{Mg}{1} lines (5172 and 5183~{\AA}) are useful for this purpose:
all other lines are too weak, or show blends with CH or Balmer lines. The FWHMs
of these two lines were measured by fitting a Gaussian profile for {\FAC} and
{\GG}, and are given together with equivalent widths in Table~\ref{tab:lw}.

Line broadening by macroturbulence and/or rotation, along with the
instrumental broadening, were also estimated by using synthetic
spectra. For these measurements, we calculated synthetic spectra of
the Mg lines using model atmospheres having the stellar parameters
determined in the abundance analysis (see the next section), and
searched for the $\chi^{2}$ minimum in the fitting of synthetic
spectra to the observed one. A Gaussian profile was assumed for the
broadening, including instrumental effects. Although the rotation and
macroturbulence produce profiles that differ from a Gaussian, this
approach provides a first approximation to line broadening by these
external effects. The free parameters in the procedure are the Mg
abundance, wavelength shift, and the Gaussian broadening. Results are
given in Table~\ref{tab:lw}, where the values corrected for the
instrumental broadening are presented. For {\FAC}, two possible
solutions of the stellar parameters (``dwarf" and ``subgiant" cases)
were derived (see \S~\ref{sec:gravity}). The difference between the
two cases is negligible ($<0.1$~km~s$^{-1}$). While the FWHMs of the
lines in {\GG} are larger than those in {\FAC}, the Gaussian
broadening is smaller in {\GG} because of the difference of the line
strengths: the equivalent widths of the lines in {\GG} are
approximately three times larger than those in {\FAC}.  The result of
these measurements is discussed in \S~5.5, comparing with the line
widths of other extremely metal-poor, main-sequence stars.

\subsection{Radial velocity measurements}

Heliocentric radial velocities were measured for the high-resolution spectra
obtained at four different epochs. Since the number of atomic lines detected
in {\FAC} is quite limited, the measurements are based only on several
\ion{Mg}{1} lines at 3830~{\AA} and 5170~{\AA}. Results are given in
Table~\ref{Tab:SubaruObs}. Measurements were also made of G~64--12 for
comparison, using the same absorption features as for {\FAC}. Although the
random error of these measurements cannot be evaluated because of the small
number of lines, a typical error for similar quality spectra using the same
technique is 0.2--0.3~km~s$^{-1}$ (e.g. Aoki et al. 2005). Assuming a
possible systematic error due to the instability of the instrument to be
0.5~km~s$^{-1}$, the total measurement error is 0.8--1.0~km~s$^{-1}$.

No clear variation of the heliocentric radial velocity was found for
{\FAC} from May 2004 to June 2005. The value derived from the low
resolution spectrum of the ESO 3.6~m telescope is $v_{\rm r}=82.6\pm
16.1$~km~s$^{-1}$, while the medium resolution spectrum of the SSO
2.3~m telescope provides $v_{\rm r}=69.5\pm 6.6$~km~s$^{-1}$.  These
agree with the results from the Subaru spectra (63.8~km~s$^{-1}$)
within the errors, although the constraint on binarity is not strong.
Further monitoring of radial velocity is required to investigate this
question.

{\GG} also shows no variation of its heliocentric radial
velocity. This has been measured by \citet{latham02} for 33 epochs
covering 13 years, and shows no clear variation. Our value
(443.7~km~s$^{-1}$) agrees, within the errors, with the average of
their measurements (442.51~km~s$^{-1}$).

\subsection{Interstellar absorption}\label{sec:ism}

\begin{figure*}[phtb]
  \includegraphics[width=10cm]{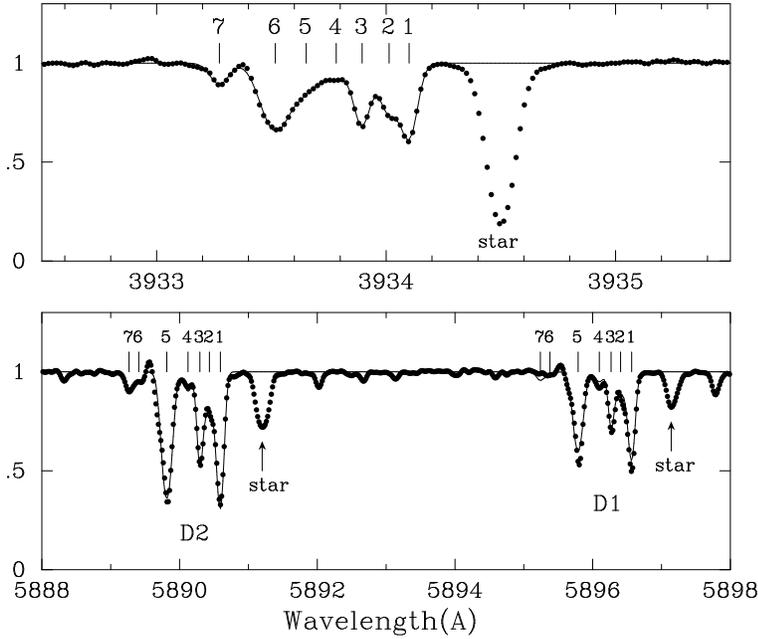}
  \caption{\label{fig:vpfit} Interstellar absorption features of the
  \ion{Ca}{2} K (upper panel) and \ion{Na}{1} D lines in the spectrum
  of {\FAC}. The wavelength is given for the heliocentric scale. Dots
  indicate the observed spectrum, while lines are those calculated
  using VPFIT (see text). The seven components derived from the
  analysis are presented for each feature.}
\end{figure*}

Significant interstellar absorption features arising from Ca and Na
are found in the spectra of {\FAC}. Fig. \ref{fig:vpfit} shows the
features of the \ion{Ca}{2} K and \ion{Na}{1} D lines. The VPFIT code
\citep{VPFIT} was used to derive the column density ($N$), radial
velocity ($v$), and Doppler parameter ($b$). We note that the fit was
made simultaneously for both \ion{Na}{1} D1 and D2 lines
($5895.9$\,{\AA} and $5889.9$\,{\AA}, respectively).

The results are presented in Tables \ref{Tab:ISCa} and
\ref{Tab:ISNa}. The radial velocity structure of the interstellar
components 1 -- 4 of both elements agrees well with each other. The
remaining components deviate because the components 5 -- 7 of
\ion{Na}{1} D lines seem to be affected by a terrestrial \ion{Na}{1} D
emission. We note that, in Fig. \ref{fig:vpfit}, the components 1, 3,
and 5 of the \ion{Na}{1} D1 line show a slight discrepancy at the core
where the fit appears less deep compared to the observation. This may
be due to a slight saturation of these components of the D2 line.

The study of interstellar features in our data provides a chance to
derive constraints on the distance of the star. According to Figure 3
of \citet{hobbs74}, the lower limit on the distance of {\FAC} might be
estimated to be about 500~pc, based on the Na column density listed in
Table \ref{Tab:ISNa} (the constraint from the Ca absorption feature is
weaker). This result should be taken into consideration in the
determination of gravity of this object (see \S \ref{sec:gravity}).

\subsection{Photometry}

$UBVRI$ photometry of {\FAC} and {\GG} was obtained with the 2\,m
MAGNUM telescope \citep{Yoshiietal:2003} on 2004 June 23 and 25 (UTC).
Both stars were observed at similar airmass ($URI$) or observed twice
to derive the airmass gradients ($BV$). The transformation of the
instrumental magnitudes for {\FAC} to values on the
Johnson-Kron-Cousins system was made differentially with respect to
{\GG}, by adopting the $UBVRI$ magnitudes provided by
\cite{Landolt:1992}\footnote{Note that {\GG} is identical to
SA105--815.} for the latter.  We repeat these values and their errors
in Table \ref{Tab:Photometry}, where we also list the results for
{\FAC}, together with $JHK$ photometry from the Two Micron All Sky
Survey (2MASS)\footnote{2MASS is a joint project of the University of
Massachusetts and the Infrared Processing and Analysis
Center/California Institute of Technology, funded by the National
Aeronautics and Space Administration and the U.S. National Science
Foundation.} All-Sky Catalog of Point Sources \citep{Cutrieetal:2003}.

$BVR$ photometry of HE 1327-2326 was also obtained with the WIYN 3.5m
telescope using the OPTIC detector \citep{howell03} on the photometric
night of June 8, 2004.  Landolt standards were taken immediately
before and after the object data, bracketing both in UT and airmass.
Aperture photometry was performed on the object and all standard
stars.  After examining the data, it was decided to use only those
standard stars that appeared in the same quadrand (and amplifier) as
the target star, thus avoiding any possible systematic calibration
errors.  This left us with a total of (9, 13, 14) standard stars for
($B, V, R$) respectively, and left us with RMS fits to the standards
that were less than 0.02 in all cases.  Unfortunately, the blue
standard that we targeted did not fall on the correct amplifier.  The
bluest remaining standard has $B-V$ of about 0.5, which is slightly
redder even than our object. The results are given in
Table~\ref{Tab:Photometry}.

$UBVRI$ photometry of the stars {\GG} and {\FAC} was also
obtained, during the night of 2005, May 22 (UTC), using the CTIO 0.9~m
telescope. Landolt standards were obtained during the course of the
night, bracketing the range of colors of these two stars. Aperture
photometry was performed on the program stars and the standards, and
final solutions for the calibrations were carried out in the usual
manner. Additional details of the observing procedures can be found
in Beers et al. (2005, in preparation). The results are also given in
Table~\ref{Tab:Photometry}.

The agreement between the MAGNUM and WIYN measurements is excellent.
While the agreement of $U$ and $V$ values between MAGNUM and CTIO
measurements is excellent, small discrepancies ($\lesssim 0.05$) are
found for others. However, if the values of {\FAC} are determined
differentially with respect to {\GG} ones, as are done for MAGNUM
data, the agreement with MAGNUM results is also excellent for all
bands. We adopted the MAGNUM and 2MASS photometry in the determination
of effective temperatures (\S \ref{sec:Teff}).

\section{STELLAR PARAMETERS}

\subsection{Effective temperature}\label{sec:Teff}

The effective temperature ($T_{\rm eff}$) of {\FAC} was estimated with
three different methods. A summary of the results can be found in
Table \ref{Tab:TeffDerivation}.

\begin{figure*}[phtb]
  \includegraphics[width=9cm]{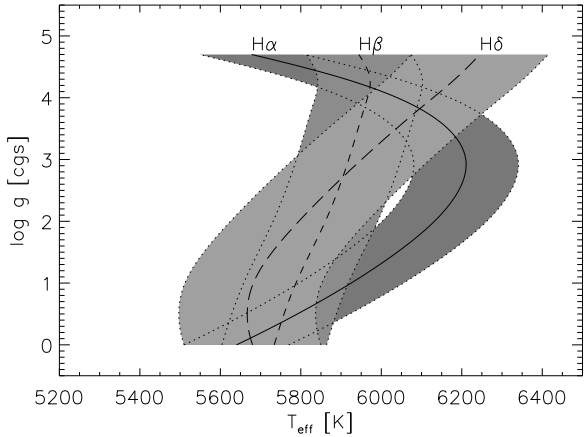}
  \caption{\label{fig:balmer} The estimated effective temperature of
  HE1327-2326 from the Balmer lines as a function of adopted $\log g$.
  The results from H$\alpha$, H$\beta$ and H$\gamma$ are shown by the
  full, short-dashed and long-dashed lines respectively.  The
  estimated error for each cases is shown by the parallel dotted lines
  and the shaded regions.}
\end{figure*}

First, we employed profile analysis of H$\alpha$--H$\delta$, following
\citet{Barklemetal:2002}.  Mixing-length parameters $\alpha = 0.5$ and
$y=0.5$ were employed. Fig. \ref{fig:balmer} shows the results of the
analyses for H$\alpha$, H$\beta$, and H$\delta$. Although two
solutions at $\log g \sim 3.4$ and 0.4 are found from this
figure\footnote{In this paper values of $\log g$ are given in cgs
units.}, the case of lower gravity is excluded for the reasons
described in the next subsection. The results are given in Table
\ref{Tab:TeffDerivation}. We note that a significant theoretical
uncertainty arises from the treatment of convection in the model
atmospheres, as can be seen from the fact that, for example, choosing
$\alpha = 1.5$ and $y=0.076$ yields 80\,K higher temperatures for
{\FAC}.

Secondly, we made use of another purely spectroscopic {\tefft}
indicator, the H$\delta$ index \texttt{HP2}. The calibration by 
\citet{Ryanetal:1999} suggests $\teffm = 6200$\,K for {\GG}, and
$6000$\,K for {\FAC}. A calibration of the \texttt{HP2} index (Beers
et al. 2005, in preparation) which is tied to the scale of
\citet{Alonsoetal:1996} yields temperatures higher by $150$\,K and
$160$\,K, respectively.

Thirdly, we derived effective temperatures for {\FAC} and {\GG} from
broadband photometry. Three independent reddening estimates are
available for {\FAC}: the maps of \citet{Burstein/Heiles:1982} yield
$E(B-V)=0.06$; those of \citet{Schlegeletal:1998} yield $0.077$. (The
maps of \citet{Schlegeletal:1998} yield $E(B-V)=0.028$ for {\GG},
while those of \citet{Burstein/Heiles:1982} yield 0.00.)  Finally,
\citet{Munari/Zwitter:1997} provide empirical relations between the
strength of the interstellar absorption of \ion{Na}{1}~D2 and
$E(B-V)$. The equivalent width of each component of the \ion{Na}{1} D2
line by interstellar medium (ISM) was measured as reported in
\S~\ref{sec:ism}, and the corresponding $E(B-V)$ was estimated, using
this empirical relations, to obtain the total value of $E(B-V)=0.104$
for {\FAC}. The equivalent width of a single component of the
\ion{Na}{1} D2 line was measured to be 19.2~m{\AA} for {\GG}, which
yields $E(B-V)=0.006$.  We adopt the Schlegel et al. reddening. The
value for {\FAC} differs by only $0.003$\,mag from the average of the
three available estimates.  The weak interstellar \ion{Na}{1} D2 line
in {\GG} suggests that the adopted reddening for {\GG} is possibly
overestimated.


We adopted the color-{\tefft} relations of \citet{Alonsoetal:1996},
which are given for Johnson $VRI$ filters and the Telescopio Carlos
Sanchez (TCS) $K$ filter. In order to apply their relation to our
study, our $VRI$ measurements on the Johnson-Kron-Cousins system and
the $K$ magnitude from 2MASS were transformed to their system.

The transformations of \citet{Bessell:1983} were used to convert the
Johnson-Kron-Cousins $V-R$ and $V-I$ colors to the Johnson system,
while dereddened 2MASS $J-K_S$ colors and equation (1c) of
\citet{Ramirez/Melendez:2004} were used to transform the 2MASS $K_S$
magnitudes to TCS $K$ values. Values of the four colour indices are
listed in the second and third columns of Table
\ref{Tab:TeffDerivation}.

We note that for stars with significant reddening, it is important to
deredden the observed magnitudes before applying the transformations, and not
vice versa. This is necessary because transformations between photometric
systems provide estimates of ratios of integrated fluxes in wavelength ranges
that are (partly) unobserved, based on ratios of observed integrated
fluxes. The transformations have been established by means of spectral energy
distributions of nearby, (almost) unreddened stars \citep[see
e.g.][]{Bessell:1983}. Since reddening affects the spectral energy
distribution of a star differently from a change of {\tefft} that would lead
to the same observed color, the transformation is valid only for one
particular reddening -- the value for the stars that were used to establish
the transformation (i.e. $E(B-V)\sim 0$). In the case of $V-R$ for {\FAC},
changing the order of dereddening and applying the relevant transformation
leads to a difference of 0.018\,mag, and a {\tefft} which is 28\,K
lower\footnote{The $R-I$ color provides a temperature estimate
that is 154\,K lower. We have chosen not to employ this color in our estimate
of {\tefft}.}. Although this result is
significantly smaller than the errors introduced by the uncertainty of the
reddening, it is systematic and should therefore be avoided.

\citet{Ryanetal:1999} note that the empirical color-{\tefft} relations
provided by \citet{Alonsoetal:1996} show an unphysical metallicity
dependence at low [Fe/H]. In their Figure 5 it can be seen that the
effect becomes significant at $\mbox{[Fe/H]}< -3.0$. Therefore, we use
their scales for $\mbox{[Fe/H]}= -3.0$ in our estimates of {\tefft}
for {\FAC} and {\GG}.

The {\tefft}--color calibrations of \citet{Houdasheltetal:2000}
provide 6390~K and 6590~K (averages) for {\FAC} and {\GG},
respectively, which are systematically higher than those derived from
\citet{Alonsoetal:1996}. This trend was also found by
\citet{Cohenetal:2002} for their dwarf and subgiant stars. We note
that the {\tefft} of {\GG} is outside the range of the
\citet{Houdasheltetal:2000} calibration (4000--6500~K). Given that
the \citet{Houdasheltetal:2000} {\tefft} scale is better calibrated
for relatively cooler stars, we adopt here the calibration of
\citet{Alonsoetal:1996}.


The averages of the effective temperatures derived from the four
above-mentioned colors yield $\teffm = 6180$\,K for {\FAC} and
$6430$\,K for {\GG}. The uncertainty of the reddening of {\FAC} of
about 0.02\,mag dominates the error in {\tefft}; it is $\sim
80$\,K. However, the excellent agreement of the above mentioned
effective temperature with $\teffm = 6160$\,K determined with the
\texttt{HP2} index confirms the reddening adopted for this star. The
slightly higher {\tefft} derived from colors for {\GG} than from the
\texttt{HP2} index suggests, again, that its reddening might have been
slightly overestimated. We adopted the average of the values derived
from the HP2 index and colors for {\GG}. The resulting values are
given in Table~\ref{Tab:StellarParameters}. It should be noted that
the difference of {\tefft} between {\FAC} and {\GG} estimated by any
method in Table \ref{Tab:TeffDerivation} is 200--300~K. Although there
remain some uncertainties in the determination of {\tefft}, the
abundance results of {\FAC} relative to those of {\GG} are very
robust.

\subsection{Surface gravity}\label{sec:gravity}

The small number of absorption lines detected in the spectrum of
{\FAC} makes the determination of its surface gravity a challenge. In
particular, the absence of \ion{Fe}{2} lines prevents the use of the
\ion{Fe}{1}/\ion{Fe}{2} ionization equilibrium. One \ion{Ca}{1} line
and two \ion{Ca}{2} lines are detected, but the Ca abundances computed
with the lines of these two species disagree by $0.57$\,dex for $\log
g=3.7$, and $0.53$\,dex for $\log g=4.5$ (see below), so that the
ionization equilibrium could only be reached for unreasonably high
surface gravities. A discrepancy of the Ca abundances from the \ion{Ca}{1}
and \ion{Ca}{2} of the same order was found for {\CBB}
\citep{HE0107_ApJ}. We consider it likely that this is caused by
non-LTE effects operating on both species \citep{korn05}.

A strong constraint on $\log g$ comes from the proper motion of {\FAC}, which
is listed in the third Yale/San Juan Southern Proper Motion Catalog
\citep[SPM3; ][]{girard04} as entry \#4266486 with
$\mu_{\alpha}=-0.0575$\,arcsec\,yr$^{-1}$ and
$\mu_{\delta}=+0.0454$\,arcsec\,yr$^{-1}$. From the requirement that the
transverse velocity of the star must not be larger than the Galactic escape
velocity, which we assume to be 500\,km\,s$^{-1}$, it follows that $M_V >
2.7$, resulting in $\log g>3.5$. Inspection of a 12\,Gyr isochrone for
$\mbox{[Fe/H]}=-3.5$ \citep{Kimetal:2002} yields that {\FAC} is either a
subgiant ($\log g=3.7$), or a main sequence star ($\log g=4.5$).

{From} the colors listed in \citet{Houdasheltetal:2000}, it may be seen that at
$\teffm = 6500$\,K and $\mbox{[Fe/H]}=-3.0$, the $U-B$ color changes with
$\log g$, and can therefore be used as a gravity indicator. In the relevant
parameter range, $\Delta(\log g)/\Delta( U-B) = -0.067\,\mbox{dex}/0.01\,
\mbox{mag}$.  Since {\FAC} is 0.047\,mag bluer in $U-B$ than {\GG}, the
gravity of the former star should be $\sim 0.3$\,dex higher.  That is, it
should be located farther down the main sequence. Although the effective
temperature scale of \citet{Houdasheltetal:2000} was not adopted in the
present study the estimate of the gravity differences between {\FAC} and
{\GG} obtained from their relations is still meaningful.

On the other hand, the relative strengths of the Balmer lines may also
be used to constrain $\log g$ \citep[see][for a description of the
method]{HE0107_ApJ}. The best agreement among the effective
temperatures derived from H$\alpha$--H$\delta$ is reached at $\log g =
3.4$ and $0.4$, respectively (Fig.~\ref{fig:balmer}). As mentioned
above, the low gravity solution is ruled out by the measured proper
motion, leaving the subgiant solution. Apart from the better agreement
among the results from the individual Balmer lines, the profiles match
the observations better if a subgiant gravity is assumed, rather than
the main sequence value of $\log g = 4.5$.

Finally, we attempted to use survey volume and detection probability
arguments to decide whether {\FAC} is a subgiant or a main sequence
star. {From} the isochrones of \citet{Kimetal:2002}, for [Fe/H]$ =
-3.5$ and age = 12 Gyr we read that the absolute visual magnitudes for
a star with $\teffm = 6150$\,K (and $(B-V) = 0.40$) on the main
sequence and on the subgiant branch are $M_V=5.1$ and $2.8$,
respectively. It follows that the survey volume for subgiants with
$\teffm = 6150$\,K in a flux-limited survey is 24 times larger than
for main sequence stars of the same {\tefft}.  One may also use the
data of \citet{Kimetal:2002} to determine the relative numbers of
subgiants and dwarfs expected in a {\it complete} sample: assuming
their mass function defined by x = 1.35 (i.e. Salpeter) we find that
for $\teffm = 6150\,K$ the expected ratio of dwarfs to subgiants is
$\sim$30.  Combination of the relative number of stars in the subgiant
and main sequence evolutionary state with the ratio of survey volumes
then yields that the probabilities of {\FAC} being a dwarf or a
subgiant are roughly equal.

We have thus carried out our abundance analysis for both gravities
inferred from the isochrones of \citet{Kimetal:2002}: $\log g = 3.7$
and $\log g = 4.5$. As shown below, the two assumptions yield very
similar results for the chemical abundances. We note that the
distances estimated from the above gravities are 1.4~kpc (subgiant
case) and 0.47~kpc (dwarf case). Neither possibility can be
excluded by the constraint from the interstellar absorption.

Recent studies for {\GG} yield a gravity of $\log g \gtrsim 4.0$
\citep[e.g. ][]{akerman04}. Assuming it to lie on the main sequence,
we estimated the gravity to be $\log g=4.38$ from the above
isochrone. The LTE iron abundance derived from \ion{Fe}{2} lines
adopting this gravity is slightly higher than that from \ion{Fe}{1}
lines (see next section and Table~\ref{Tab:Abundances}). This
discrepancy could, however, be explained by NLTE effects for the
\ion{Fe}{1} lines \citep[e.g. ][]{thevenin99,Kornetal:2003}, rather
than errors in the gravity estimate.

\subsection{Microturbulence}

Apart from the lines of hydrogen, the only strong features in the
spectrum of {\FAC} are the H and K lines of \ion{Ca}{2}.  However, no
weak \ion{Ca}{2} line is detected, and it is therefore impossible to
adjust the microturbulent velocity {\vmicrt} in the usual way by
requiring that no trend of the abundance of a given species is seen
with line strength. That said, given the general line weakness, the
choice of {\vmicrt} does not significantly influence the results of
our abundance analysis for all but one species.  The exception, of
course, is \ion{Ca}{2}, the abundance of which exhibits a strong
dependence on {\vmicrt} of almost 0.4\,dex per km~s$^{-1}$. Therefore,
well-justified values for {\vmicrt} should be chosen for the analysis
of this species.

We determine {\vmicrt} empirically by averaging over values obtained
by \citet{Cohenetal:2004} for very metal-poor subgiants and main
sequence stars with effective temperatures similar to that of
{\FAC}. In selecting stars for this exercise, we take into account
that \citet{Cohenetal:2004} adopt the {\tefft} scale of
\citet{Houdasheltetal:2000}, which is $\sim 200$\,K higher than the
scale we adopt here, and accept stars with $6280-6480$~K from their
sample. We thus obtain \vmicrt$ = 1.7$\,km~s$^{-1}$ for $\log g =
3.7$, and $\vmicrm = 1.5$\,km~s$^{-1}$ for $\log g = 4.5$.

The value of {\vmicrt} for {\GG} was determined from 59 \ion{Fe}{1}
lines by requiring no dependence of the derived abundances on line
strengths. The result is 1.6~km~s$^{-1}$. This value agrees very well
with that adopted for {\FAC}.

\section{ABUNDANCE ANALYSIS}\label{sec:ana}

In this section we report on our model atmosphere abundance analyses
for {\FAC} and {\GG}. Table~\ref{Tab:Abundances} provides the
abundances derived by W.A. using model atmospheres of
\citet{kurucz93}, along with those obtained by A.F.  using MARCS
models (see below) reported in \citet{frebel05}. The agreement between
the analyses is fairly good in general. We first compare the model
atmospheres of the two grids (\S~\ref{sec:model}). In
\S\S~\ref{sec:iron}--\ref{sec:li} the details of the abundance
analyses and results, along with error estimates, are
discussed. Abundances are determined for gravities corresponding to
the main sequence and subgiant cases discussed above.

\subsection{Model atmospheres}~\label{sec:model}

\begin{figure*}[phtb] 
  \includegraphics[width=9cm]{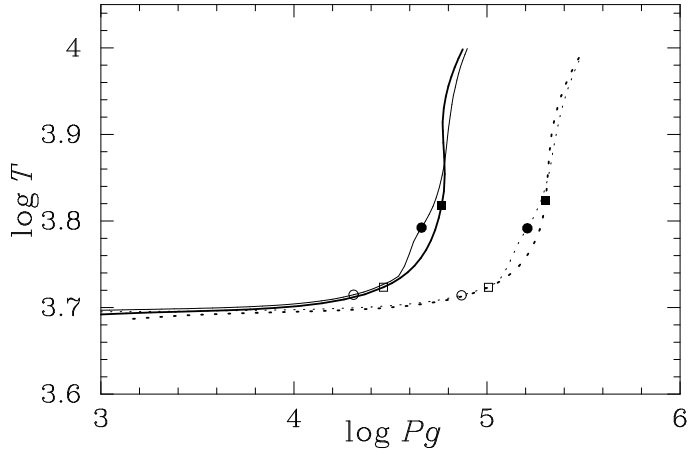}
  \caption{\label{fig:model} Thermal structures of MARCS (thick lines)
  and Kurucz (thin lines) model atmospheres for $\log g=3.7$
  (solid lines) and 4.5 (dashed lines). Filled and open circles
  indicate the points where the optical depth at 5000~{\AA} is 1.0 and
  0.1, respectively, in the Kurucz models, while filled and open
  squares indicate those in the MARCS models.}
\end{figure*}

Fig.~\ref{fig:model} shows the thermal structures of model atmospheres
used in the analysis. In this figure, the temperature is shown as a
function of the gas pressure, and the points where the optical depth
at 5000~{\AA} is 1.0 and 0.1 are presented.
We obtained the 'Kurucz' models for $T_{\rm
eff}=6180$~K and $\log g=3.7$ and 4.5 by interpolating the grid for
[Fe/H]$=-5.0$, which was calculated using an overshooting approximation
\citep{castelli97} for the scaled solar abundances. The MARCS models
\citep[e.g. ][]{gustafsson75} were calculated using the latest version
of the code (Gustafsson et al., in preparation) for [Fe/H]$=-5.0$
assuming the enhancements of C ([C/Fe]=+2.0) and alpha elements
([$\alpha$/Fe]=+0.5). Comparison of the models in Fig.~\ref{fig:model}
depicts that the thermal structures agree well in general, even though
the assumed carbon abundances are significantly different. Since
carbon-bearing molecules like CH and CN are not strongly represented
in this object because of its high temperature ({\tefft}$=6180$~K),
enhanced carbon and nitrogen do not provide important sources of
opacity. The difference in the thermal structure in internal regions
($\log T\gtrsim 3.8$) would be due to the differences in the treatment
of convection between the two model calculations.

\subsection{Iron}\label{sec:iron}

Seven \ion{Fe}{1} lines are seen in our spectrum of {\FAC}. Three of
them are blended with Balmer lines, and only four were used to
determine the final Fe abundance of this object. The two lines at
3581.2\,{\AA} and 3859.9\,{\AA} are clearly detected (see
Fig.~\ref{Fig:FeLines} for an illustration); they have strengths in
the range 5--7\,m{\AA}. Two further lines at 4045.8\,{\AA} and
3820.4\,{\AA} are barely detected. Using only these four lines we
derive iron abundances of $\mbox{[Fe/H]}=-5.66$ and
$\mbox{[Fe/H]}=-5.65$ for the subgiant and main sequence gravities,
respectively.

The equivalent widths of the other three lines (3737.1, 3745.6, and
3758.2~{\AA}), given in Table~\ref{Tab:Eqw}, were measured by regarding
the wing profile of Balmer lines as a pseudo-continuum. This simple
assumption would be justified by the fact that the higher order Balmer
lines are formed in very deep atmospheric layers, while the
\ion{Fe}{1} lines with low excitation potential are formed in
relatively higher layers. The result of the analysis including these
three lines (i.e. using all seven lines detected) gives 
an iron abundance higher by 0.05~dex than the result from the four lines. This
difference is smaller than the random error of Fe abundance (see
below). We note that a spectrum synthesis for Fe and Balmer lines was
not attempted, because of the difficulty in estimating the real
continuum.

As a check of the reality of the \ion{Fe}{1} line detections, we
computed predicted strengths for an extensive list of \ion{Fe}{1}
lines for $\mbox{[Fe/H]}=-5.5$. Though the line at 3719.9\,{\AA} is
predicted to be strong enough to be detectable, this line is not
clearly detected because the spectrum is severely disturbed by a
cosmic ray hit at this wavelength.

Numerous CH lines exist in the spectrum of {\FAC}, as seen in our
Fig.~\ref{Fig:Gbandfit} as well as in Figure 1 of
\citet{frebel05}. It is important to note that our spectrum synthesis
calculations, which include CH lines, confirmed that none of the
\ion{Fe}{1} lines we use are blended with CH lines of measurable
strength.

The random error in the Fe abundance was estimated for 1$\sigma$
errors of the equivalent width measurements, determined using the
formula of \citet{Norrisetal:2001} for individual \ion{Fe}{1}
lines. The derived value is given in Table~\ref{tab:error}. Errors in
[Fe/H] due to uncertainties in the atmospheric parameters were
estimated for changes of $\sigma (T_{\rm eff})=100$~K, $\sigma (\log
g)=0.3$~dex, and $\sigma (v_{\rm micr})=0.3$~km~s$^{-1}$, as given in
Table~\ref{tab:error}. The root-sum-square value of these four sources
is adopted as the final abundance error. The random error of the
observational measurement and that arising from the effective
temperature estimate are the dominant sources of uncertainty in the Fe
abundance determination from \ion{Fe}{1} lines.

We adopt $\Delta \log \epsilon$(Fe)$=+0.2$~dex as the NLTE correction
for the iron abundance determined from \ion{Fe}{1} lines, based on the
studies by \citet{Kornetal:2003}, \citet{Grattonetal:1999},
\citet{thevenin99}, and Collet \& Asplund (2005, in preparation), as
discussed by \citet{asplund05}. 

A significant 3D effect on the formation of Fe lines is also
expected. Calculations including this effect results in lower Fe
abundance, and the effect is significant in particular in metal-poor
stars \citep{asplund99}. However, the direction of the correction for
this effect is opposite to that for an NLTE one. A full analysis
including both 3D and NLTE effects is required to estimate the
uncertainties of our analysis by neglecting these effects.

The upper limit of the Fe abundance estimated from the \ion{Fe}{2}
5018.5~{\AA} line is one order of magnitude higher than the Fe
abundance determined from neutral lines. This upper limit is too high
to usefully constrain the gravity from the \ion{Fe}{1}/\ion{Fe}{2}
ionization balance.

\subsection{C, N, and O} 

\begin{figure}[htbp]
  \includegraphics[width=9cm]{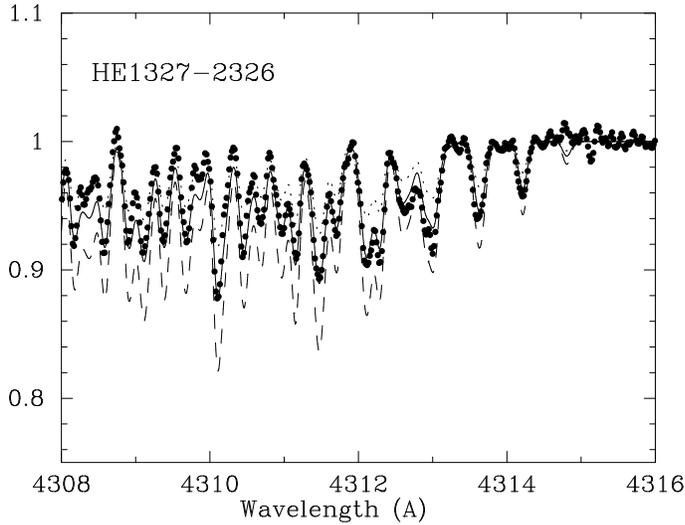}
  \caption{\label{Fig:Gbandfit} Synthetic spectra of the CH G-band
  (lines) and the observed spectrum of {\FAC} (dots). The assumed
  carbon abundance ratios in the calculations are [C/H]$=-1.6$ (dotted
  line), $-1.4$ (solid line), and $-1.2$ (dashed line).}

\end{figure}

The carbon abundance was measured with spectrum synthesis calculations
of weak lines of the {\it A--X} and {\it B--X} systems of CH, as well
as the G band at $4310$\,{\AA} (see Fig.~\ref{Fig:Gbandfit}). The line
list of this molecule was produced using molecular constants provided
by \citet{zachwieja95} and \citet{kepa96}, and $f$-values of
\citet{brown87}. The spectrum synthesis using this list and the solar
model atmosphere by \citet{Holweger/Mueller:1974} reproduces the solar
carbon abundance ($\log\epsilon$(C)=8.55) derived by
\citet{grevesse96} . In this paper, however, we adopt the solar carbon
abundance ($\log\epsilon$(C)=8.39) determined by
\citet{asplund05c}. The $\log\epsilon$(C) values in
Table~\ref{Tab:Abundances} are the carbon abundances corrected for
this revision of the solar abundance. The C abundances determined from
the different features of CH agree to better than $0.1$\,dex. The {\it
C--X} system of CH at 3140~{\AA} is also detected, but is not used to
determine the C abundance because of the low quality of the spectrum
at this wavelength.

\begin{figure}[htbp]
  \includegraphics[width=9cm]{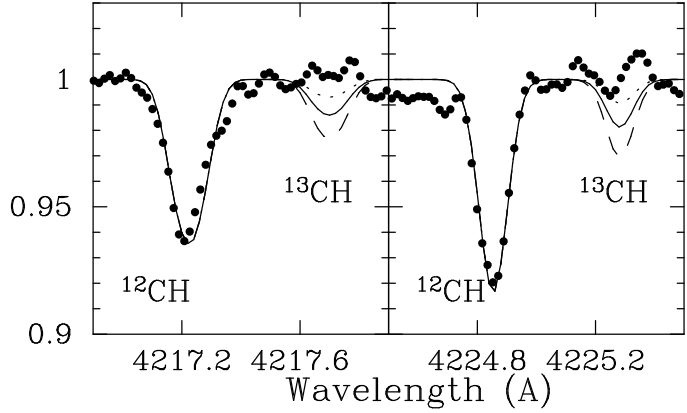}
  \caption{\label{fig:ciso} Same as Fig.~\ref{Fig:Gbandfit},
but for $^{12}$CH and $^{13}$CH molecular features. The assumed
$^{12}$C/$^{13}$C ratios are 3 (dashed line), 5 (solid line), and 8
(dotted line). From the comparison with the observed spectrum (filled
circles), the lower limit of $^{12}$C/$^{13}$C is estimated to be 5.}
\end{figure}

No $^{13}$CH line is detected in the {\FAC} spectrum. The limit on the carbon
isotope ratio ($^{12}$C/$^{13}$C) was estimated from the CH molecular
features around 4220~{\AA}, where the line positions of the $^{13}$CH $A-X$
band were calculated using the molecular constants of
\citet{zachwieja97}. Fig.~\ref{fig:ciso} compares the observed spectrum
with synthetic ones calculated assuming $^{12}$C/$^{13}$C=3, 5, and 10.  {From}
this comparison we estimate the lower limit of the $^{12}$C/$^{13}$C to be
5. This is a $3\sigma$ limit if a S/N ratio of 200/1 is adopted. In order to
obtain a tighter lower limit of this isotope ratio, a higher quality spectrum
is needed. The lower limit estimated in the present analysis, however,
is already higher than the equilibrium value of the CNO cycle
($^{12}$C/$^{13}$C=3--4). This could be an important constraint on models
seeking to explain the origin of the large enhancement of C and N in this
object.

\begin{figure}[htbp]
  \includegraphics[width=9cm]{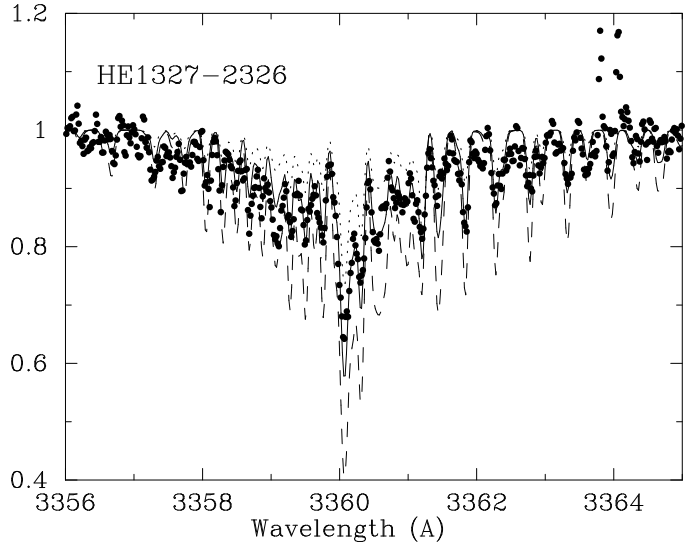}

  \caption{\label{Fig:NHfit} Same as Fig.~\ref{Fig:Gbandfit}, but for
  the NH 3360~{\AA} band. The assumed nitrogen abundances in the
  calculation are [N/H]$=-1.4$ (dotted line), $-1.1$ (solid line), and
  $-0.8$ (dashed line).}

\end{figure}

The NH band at $3360$\,{\AA} is detected in our spectrum
(Fig.~\ref{Fig:NHfit}). We first calculated the synthetic spectrum of this
feature using the line list provided by \citet{kurucz93n15} and the solar
model \citep{Holweger/Mueller:1974}, and found that to fit the solar
spectrum requires a nitrogen abundance lower by $\sim$0.4~dex
than that of \citet{asplund05}. We therefore systematically reduced
the $\log gf$ values in Kurucz's NH line lists by 0.4~dex in the present
work.\footnote{We also attempted the analysis using an NH line list kindly
provided by A. Ecuvillon that was used in previous work
\citep{ecuvillon04}. We found the solar spectrum of the NH band is well
reproduced by the line data, while the feature of {\FAC} was not. Further
investigation of the line data is desired for future analysis for high-quality
spectra.}

Spectrum synthesis (see Fig. \ref{Fig:NHfit}) yields $\mbox{[N/H]}=-1.1$
for $\log g = 3.7$ and $\mbox{[N/H]}=-1.6$ for $\log g = 4.5$. Unfortunately,
the uncertainty of the continuum estimate is rather large, as can be seen in
Fig.~\ref{Fig:NHfit}. This is an important source of error in the N
abundance determination.

{From} the comparison of synthetic spectra with the observed one
(Figures~\ref{Fig:Gbandfit} and \ref{Fig:NHfit}), we estimate a fitting error
of 0.1 and 0.2~dex for CH and NH bands, respectively. The systematic errors
related to atmospheric parameters, estimated by the same method as for Fe
described in \S~\ref{sec:iron}, are given in Table~\ref{tab:error}. The final
total errors of C and N abundances are 0.24 and 0.30~dex, respectively.

Significant 3D effects on CH and NH lines are expected (Asplund 2004;
Asplund et al. 2005a). The corrections for the C and N abundances from
these molecular features are estimated to be $-0.5$ -- $-1.0$~dex
depending on the lines used for the analysis, and they are not
compensated by NLTE effects. Since we adjusted the {\it gf}-values to
fit the CH and NH features in the solar spectrum, these corrections
would be partially compensated. However, the 3D effect in metal-deficient
stars is more significant than in metal-rich ones. We here point out
that systematic errors due to the 3D effect might be included in our
1D analysis, in addition to the uncertainties mentioned above.

{From} the absence of UV-OH lines, we determine an upper limit for the
oxygen abundance of $\mbox{[O/H]}<-1.66$ for the subgiant stellar
parameters, and $\mbox{[O/H]}<-1.96$ for the main sequence parameters
in \citet{frebel05}. We did not repeat the analysis for OH lines in
this work, and adopt the results of \citet{frebel05} here.

Because no clear features from the above molecules are found in our
spectrum of {\GG}, we did not attempt to measure its C, N, and O
abundances. For comparison purposes, the abundances determined by
previous studies are given in Table~\ref{Tab:Abundances}. It should be
noted that while the N abundance was estimated from the NH molecular
features by \citet{israelian04}, those of C and O were determined from
\ion{C}{1} and \ion{O}{1} lines in the near infrared by
\citet{akerman04}.  {\GG} shows no clear excess of C and O compared
with values found in other extremely metal-poor stars, while a large
overabundance of N is found.

\subsection{Elements from Na ($Z=11$) to Zn ($Z=30$)}\label{sec:na_zn}

The Na abundance of {\FAC} was measured from the \ion{Na}{1} D
lines. The contamination by telluric lines was removed by using the
spectrum of a rapidly rotating early-type star, and is not a
significant source of error. The difference in abundance derived using
different model atmospheres (Kurucz and MARCS models), is largest
(0.14 and 0.11~dex for subgiant and dwarf cases, respectively) among
the elements analyzed here. The reason for this relatively large
discrepancy is unclear.

The \ion{Na}{1} D lines are known to be severely affected by
departures from LTE. We adopted the NLTE correction of $-0.2$~dex
estimated by \citet{takeda03}. A large
overabundance of Na is found in {\FAC} ([Na/Fe]$\sim +2.1$), even
after the NLTE correction. For {\GG} [Na/Fe] has also been estimated
from the \ion{Na}{1} D lines by Aoki et al. (2005, in preparation)
using a very high S/N spectrum obtained with Subaru/HDS. The Na
abundance of this star exhibits a rather large underabundance (the
NLTE corrected value is [Na/Fe]$=-1.53$) compared with other extremely
metal-poor stars (e.g. five stars with [Fe/H]$<-3.5$ studied by Cayrel
et al. 2004 have [Na/Fe] $= -0.70$, after the NLTE correction of
$-0.5$~dex, on average).

\begin{figure}
\includegraphics[width=8.5cm]{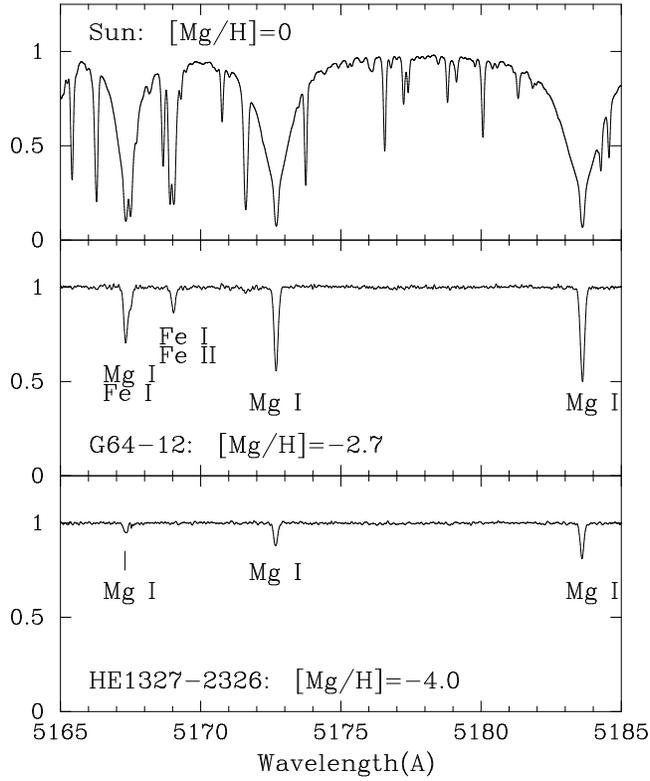}
\caption[]{Same as Fig. 1, but for the \ion{Mg}{1} b lines.}\label{Fig:Mgb} 
\end{figure}

The Mg triplets at 3820~{\AA} and 5170~{\AA} are clearly detected in
the {\FAC} spectrum (Fig.~\ref{Fig:Mgb}). Since two lines of the
former system blend with a Balmer line (H9), only the remaining four
are used in the abundance analysis. The NLTE effect is estimated not
to be large ($\sim +0.1$~dex: e.g. Gehren et al. 2004). One
sees that Mg also exhibits a large overabundance ([Mg/Fe]$\sim +1.7$)
in {\FAC}, while that of {\GG} ([Mg/Fe]$=+0.46$) is typical of values
found in other extremely metal-poor stars
\citep[e.g. ][]{Cohenetal:2004}.

The Al abundance was determined from the doublet line of \ion{Al}{1}
at 3961~{\AA}. (Although the other component of this doublet at
3944~{\AA} is clearly detected, it is blended with CH lines. Spectrum
synthesis for the 3944~{\AA} line, including CH features, provides a
consistent result with that from the standard analysis of that at
3961~{\AA}.) A significant NLTE effect is known for this \ion{Al}{1}
line ($\sim +0.6$~dex: e.g. Baum\"{u}ller et al. 1997; Gehren et
al. 2004). The Al abundance of {\GG} was also determined, yielding
[Al/Fe] = $-0.52$ (LTE), which is typical for extremely metal-poor
objects \citep[e.g. ][]{Cohenetal:2004}. The [Al/Fe] value (LTE) of {\FAC} is
approximately 1.7~dex higher than that of {\GG}. This large excess of
Al is discussed in detail in \S 5.

\begin{figure}[htbp]
  \includegraphics[width=9cm]{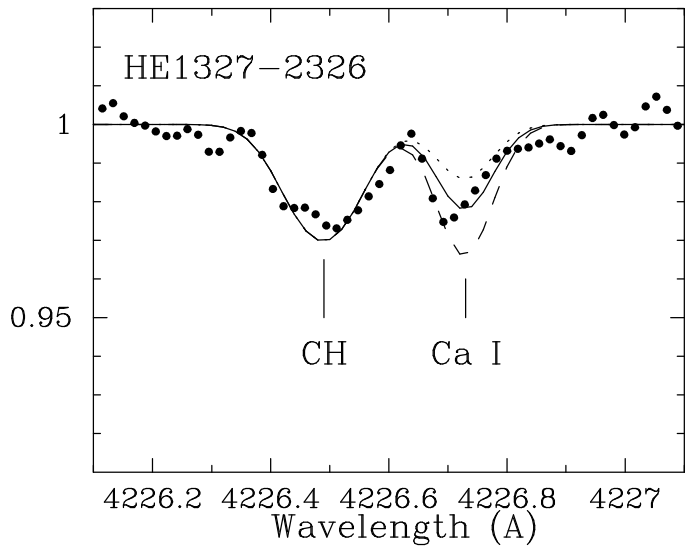}
  \caption{\label{fig:ca1} Same as Fig.~\ref{Fig:Gbandfit}, but
  for the \ion{Ca}{1} 4226~{\AA} line. CH molecular features are
  included in the calculation. Assumed Ca abundances are [Ca/H]$=-5.61$
  (dotted line), $-5.41$ (solid line), and $-5.21$ (dashed line).}
\end{figure}

The \ion{Ca}{2} K line of {\FAC} is clearly separated from the complex
interstellar absorption feature, as shown in Figure 1 of
\citet{frebel05}, while the \ion{Ca}{2} H line at 3968~{\AA} blends
with a Balmer line (H$\epsilon$). The Ca abundance derived from the K
line shows a large excess in {\FAC} ([Ca/Fe]$\sim +0.9$).  By way of
contrast, the very weak absorption of \ion{Ca}{1} 4226~{\AA}
(Fig.~\ref{fig:ca1}) yields a relatively low abundance ([Ca/Fe]$\sim
+0.3$). The large difference between the abundances derived from these
two species suggests significant NLTE effects, and further
investigation of the weak feature of \ion{Ca}{1}, as well as NLTE
effects, is required to solve this discrepancy.  We adopt the average
of the Ca abundances determined from the two species in the following
discussion.  The Ca abundance of {\GG} was estimated from six
\ion{Ca}{1} lines including the 4226~{\AA} feature.  These six lines
give a consistent result within the errors. The [Ca/Fe] value of this
star is similar to those found in other extremely metal-poor stars
\citep[e.g. ][]{Cohenetal:2004}.

The Ti abundance of {\FAC} was determined from two weak \ion{Ti}{2}
lines in the near-UV range. The relatively low quality of the spectrum
in this wavelength range leads to a rather large abundance
uncertainty: the estimated random error is 0.17~dex. Two other
\ion{Ti}{2} lines at 3759.3 and 3961.3~{\AA} are detected, though they
exist in the wings of Balmer lines. As for the \ion{Fe}{1} lines, we
applied our analysis to the equivalent widths of these lines measured
assuming the wing profile of the Balmer lines as a pseudo-continuum
for comparison purposes. The abundance derived from these two lines is
0.29~dex higher than that from the two lines in the near-UV range.
This size of discrepancy might be reasonable, given the uncertainties
of abundance determination from both line sets. The derived Ti
abundance values ([Ti/Fe]=+0.5 to +0.7) show some excess of Ti with
respect to Fe, but not as large as those of the lighter $\alpha$
elements (e.g. Mg).  In the near-UV spectrum of {\GG}, a number of
\ion{Ti}{2} lines are detected; the derived Ti abundance shows a small
excess.

The abundance errors for the above elements were estimated in the same
manner as for Fe (Table~\ref{tab:error}). The errors due to the
uncertainties in equivalent width measurements dominate the final
errors of Ca and Ti, because of the weakness of the lines and the limited
S/N of the UV spectrum.
 
The iron-group elements Cr, Mn, Co and Ni, as well as Zn, are not
detected in {\FAC}, although we attempted to search for lines which
are found in other metal-poor stars, as mentioned in
\S~\ref{sec:ew}. The 3$\sigma$ upper limits on the abundances for
these elements are given in Table~\ref{Tab:Abundances}. In general,
these upper limits, except for that of Ni, are too high to usefully
constrain any nucleosynthesis models.  The limit for Ni might have
some significance for the supernova models proposed to explain the
iron-deficient stars \citep[e.g. ][]{Umeda/Nomoto:2003}. A higher
quality UV spectrum is needed to determine the abundances of Ni and
other elements, or to provide tighter upper limits.

\subsection{Neutron-capture elements}\label{sec:ncapture}

\begin{figure}[htbp]
\includegraphics[width=8.5cm]{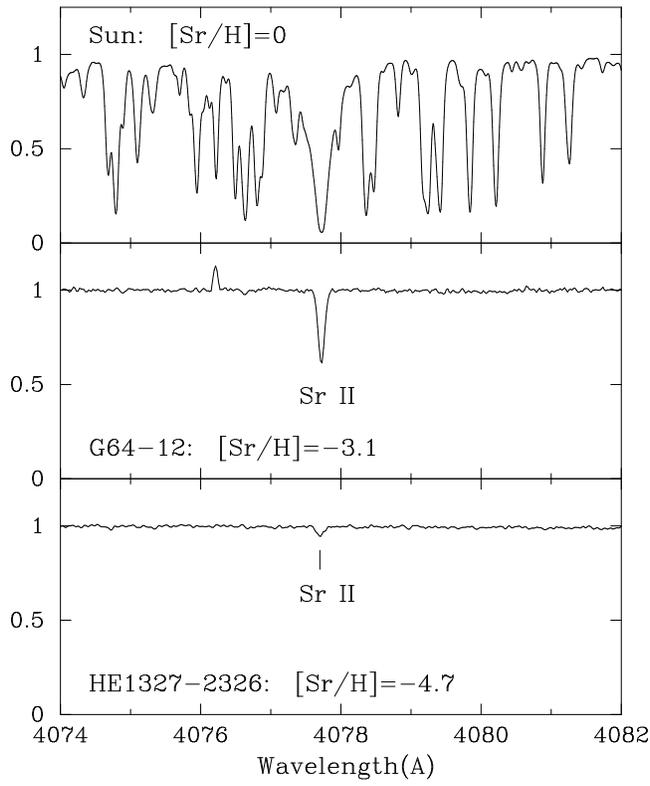}
  \caption[]{Same as Fig. 1, but for the \ion{Sr}{2}  4077~{\AA} line.}\label{Fig:Sr} 
\end{figure}

The \ion{Sr}{2} resonance lines at 4077~{\AA} and 4215~{\AA} are clearly
detected in both {\FAC} and {\GG} (see Fig.~\ref{Fig:Sr}).  Although the
\ion{Ba}{2} resonance lines are also relatively strong in extremely metal-poor
stars, including {\GG}, they are not seen in the spectrum of {\FAC}, in spite
of the spectrum having its highest $S/N$ ratio at these wavelengths. The
3$\sigma$ upper limit on the Ba abundance in {\FAC} was estimated from the
4554~{\AA} line, leading to a lower limit of the Sr/Ba ratio that plays an
important role in constraining the origin of neutron-capture elements
(\S~\ref{sec:sr}). No other neutron-capture element is detected in {\FAC},
and no useful upper limit is derived from our spectrum.

\subsection{Lithium}\label{sec:li}

\begin{figure}[htbp]
  \includegraphics[width=9cm]{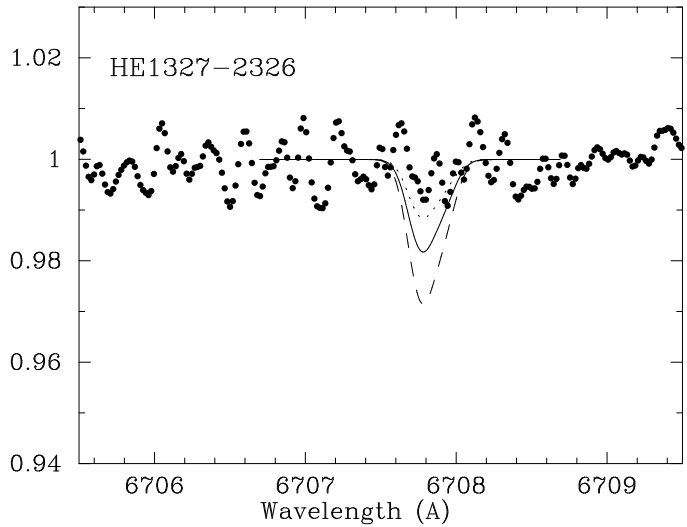}
  \caption{\label{Fig:LiSynthesis} Same as Fig.~\ref{Fig:Gbandfit},
but for the \ion{Li}{1} 6707\,{\AA} doublet. Assumed Li abundances are
$\log\epsilon\left(\mbox{Li}\right) = 1.3$, (dotted line) $1.5$ (solid
line), and $1.7$ (dashed line).}
\end{figure}

We do not detect the \ion{Li}{1} doublet at 6707\,{\AA} in {\FAC}. The
detection formula of \citet{Norrisetal:2001} yields a
$3\,\sigma$-level limit of 6.9\,m{\AA}, from which we derive an upper
limit for the Li abundance of $\log\epsilon\left(\mbox{Li}\right) <
1.5$. This can be seen in Fig. \ref{Fig:LiSynthesis}, where
comparisons of synthetic spectra adopting
$\log\epsilon\left(\mbox{Li}\right) = 1.3, 1.5,$ and 1.7,
respectively, with the observed spectrum are shown. The non-detection
of Li in this object is in strong contrast with the value expected
from the Spite plateau.


{\GG} is known to share the Li abundance of the Spite plateau. In
Table~\ref{Tab:Abundances} we provide the value determined by Aoki et
al. (2005, in preparation).


\section{Chemical abundance characteristics}

We now review the elemental abundances of {\FAC} and discuss their
implications, by comparing our results with those of other extremely
metal-poor stars. Fig.~\ref{fig:xh} shows the values of [X/H] as a
function of atomic number for {\FAC} and {\CBB}.  For this
illustration we have used the results for the subgiant case for {\FAC}
including NLTE corrections, which we have also applied to the
abundances of {\CBB} as reported by \citet{HE0107_ApJ}. For
comparison purposes the chemical abundance patterns of the average of
extremely metal-poor stars with [Fe/H]$<-3.5$, and those of the
carbon-rich objects CS~22949--037 and CS~29498--043 (Depagne et
al. 2002; Aoki et al. 2004) are shown by lines with open and filled
circles, respectively.

\subsection{Fe abundance}

The most remarkable result of this investigation is the low Fe abundance of
{\FAC}. The NLTE-corrected value of [Fe/H] $=-5.4$ is more than one order of
magnitude lower than the Fe abundances of previously known metal-poor dwarfs
and subgiants (e.g. CS~22876--032 with [Fe/H] $=-3.7$; Norris et
al. 2000). The Fe abundance of {\FAC} is even lower, by 0.2~dex, than that of
{\CBB}, a giant with an outstandingly low Fe abundance compared to that of other
low-metallicity stars. Taking into account the unavoidable errors in the
determination of Fe values in both of these stars, {\FAC} and {\CBB} should
be regarded to have similar Fe abundances, well-separated in [Fe/H] from all
other known stars.

The stars with the lowest iron abundances, other than {\FAC} and
{\CBB}, have [Fe/H]$\sim -4$
\citep[e.g. ][]{Cayreletal:2004}\footnote{The iron abundance of
CD$-38^{\circ}245$ determined by \citet{Cayreletal:2004} is
[Fe/H]$=-4.2$, i.e. slightly lower than $-4.0$
\citep[e.g. ][]{Norrisetal:2001}. This depends, however, on the
atmospheric parameters adopted in the analysis. Here, we regard the
iron abundance of this star to be [Fe/H]$\sim -4$ in order to simplify
the discussion.}. Quite recently the iron abundance of the dwarf
carbon star G~77--61 was redetermined to be [Fe/H]$\sim -4$
\citep{plez05}. Thus, there exists at least a one dex gap between the
two most iron-deficient stars ({\FAC} and {\CBB}) and all others. This
is an unexpected result, differing significantly from models of the
metallicity distribution function reported before the discovery of
{\CBB} in 2001. For instance, the model of \citet{Tsujimotoetal:1999},
which explains well the metallicity distribution function for stars
with [Fe/H]$>-4$, predicts a continuously, and more rapidly,
decreasing trend for the fraction of stars in the lower [Fe/H]
range. The existence of two stars with [Fe/H]$<-5$, along with the
absence of stars with $-5<$[Fe/H]$<-4$, suggests that some
unidentified mechanism has worked to produce the observed
iron-abundance distribution.

\subsection{C overabundance}

Another similarity between {\FAC} and {\CBB} is their large overabundance of
carbon ([C/Fe]$\sim +4$). Fig.~\ref{fig:cfe} shows the carbon abundance
ratios as a function of [Fe/H] for very metal-poor stars, including the above
two. The high [C/Fe] values together with the low Fe abundance clearly
distinguish these two stars from all others.  It should be noted, however,
that there exist two extremely metal-poor ([Fe/H]$=-4.0$ to $-3.5$) stars
with large overabundances of carbon (CS~22949--037 and CS~29498--043). The
relationship with these carbon-rich objects will be discussed in the next
subsection.

It should be noted that the C overabundance of {\FAC} was {\it not} known at the
point of sample selection. Although the overabundance of C relative to Fe is
very large, as a result of its high effective temperature molecular features are
not strong in this object. The excess of C was only recognized, for the first
time, from the high-resolution Subaru spectrum. This was also the case for
{\CBB}, the other HMP star with [Fe/H]$<-5$; the carbon richness of this star
was first appreciated only when it was observed at moderate spectral resolution.
One may therefore exclude the possibility that our recognition of the strong
carbon overabundance in either of these stars results from a selection bias.
Both were selected as metal-poor candidates based solely on the weak Ca II K
lines in their objective-prism spectra.

\subsection{Excesses of Na, Mg and Al}

Remarkable differences are found in the abundance patterns of
{\FAC} and {\CBB} for the elements Mg and Al. While the abundance
ratios of Mg and Al in {\CBB} are not high ([Mg/Fe]$=+0.15$ and
[Al/Fe]$<-0.26$, Christlieb et al. 2004) and similar to those of stars
with [Fe/H]$\gtrsim-4$, {\FAC} exhibits clear excesses (see
Table~\ref{Tab:Abundances}). Fig.~\ref{fig:xh} demonstrates that the
abundance ratios in {\FAC} gradually decrease with increasing atomic
number from C to Fe.

Fig.~\ref{fig:mgfe} shows [Mg/Fe] as a function of [Fe/H] for
extremely metal-poor stars, including these two objects. One sees that
{\FAC} shares its Mg overabundance with the carbon-enhanced,
metal-poor stars CS~22949--037 (McWilliam et al. 1995; Norris et
al. 2001; Depagne et al. 2002) and CS~29498--043
\citep{Aokietal:2002d,aoki04}, in contrast with the value for
{\CBB}. The high Mg/Fe ratios in these objects can be interpreted as a
result of nucleosynthesis by supernovae that have ejected only small
amounts of material from the vicinity of the iron core
\citep{Tsujimoto/Shigeyama:2003,Umeda/Nomoto:2003}. If {\FAC} is
related to these carbon-rich, extremely metal-poor stars, a similar
explanation might be applied to {\FAC}, as well as to {\CBB} (see
\S~\ref{sec:pop2}).

The Al/Mg ratio of {\FAC} ([Al/Mg]$\sim -0.5$, assuming LTE) is
significantly higher than the values found in other extremely
metal-poor giants ([Al/Mg]$\sim -1.0$ in LTE, e.g. Cayrel et al. 2004,
Honda et al. 2004). While measurements of Al abundances have been made
only for a small number of extremely metal-poor dwarfs, the [Al/Mg]
value of {\GG} ($-0.98$: Table~\ref{Tab:Abundances}) and of
CS~22876--032 ($-0.95$, Norris et al. 2000) are clearly lower than
that of {\FAC}. The above comparisons suggest, therefore, that the
relatively high Al/Mg ratio of this object is not the result of a
difference in NLTE effects between giants and dwarfs, in spite of the
large NLTE effects (0.6~dex) expected for Al.


The excess of Al might be related to that of another odd-numbered element,
Na. The [Na/Mg] value for {\FAC}, without NLTE correction, is $\sim +0.55$,
much higher than that of {\GG} ([Na/Mg]$=-1.6$) and those found in extremely
metal-poor giants ([Na/Mg]$\sim -1$, Cayrel et al. 2004). {\CBB} has a
similarly high Na/Mg ratio ([Na/Mg]$=+0.66$, Christlieb et al. 2004b). Na
overabundances are also found in the carbon-enhanced, extremely metal-poor stars
CS~22949--037 and CS~29498--043, which also exhibit large overabundances of Mg
and Al. Their [Na/Mg] values ($\sim -0.2$, Cayrel et al. 2004, Aoki et
al. 2004) are significantly higher than those found in other extremely metal-poor
giants, though not as high as that of {\FAC}.


\subsection{N and O abundances}\label{sec:on}

The N abundance of {\FAC} is nearly two orders of magnitude higher
than that of {\CBB}. This was determined from the NH molecular band at
3360~{\AA} for {\FAC}, while the CN molecular band at 3880~{\AA} was
used for {\CBB} \citep{HE0107_ApJ}. Although possible systematic
differences in N abundances determined from these two different
features has been advocated \citep[e.g.  ][]{spite05}, the suggested
systematic difference is small compared to the extremely large N
overabundances of these two stars.

The high N abundance of {\FAC} cannot be explained by internal
processes, including CN-cycling and mixing, since it is either a
subgiant or main-sequence object. The lower limit of the
$^{12}$C/$^{13}$C ratio ($>5$), which is higher than the equilibrium
value obtained from the CN cycle, also supports this conjecture,
though the constraint from the carbon isotope ratio is still
weak. While mixing effects might occur in the giant {\CBB}, its low
$^{13}$C content \citep{HE0107_ApJ} suggests it is also not
significantly affected by internal processes.  The large overabundance
of N in {\FAC} (and possibly in {\CBB}) thus require an extrinsic
origin. We note further that the excesses of N might be related to the
large excesses of the odd-numbered elements Na and Al.

The constraint on the O abundance ([O/Fe]$<+3.7$ to $+4.0$) is still
weak.  If the C/O ratio of {\CBB} is assumed, [O/Fe] of {\FAC} would
be +2.4. Taking the enhancement of Mg into account, however, the
expected O abundance for {\FAC} might be higher. Since the abundance
of O is key to discriminating between scenarios to explain the
abundance pattern of this object, further investigations for OH
molecular lines in the UV range are urgently required.  

\subsection{Li depletion}\label{sec:lidep}

One quite unexpected result of our investigation is the non-detection
of Li in {\FAC}. The upper limit on its Li abundance is $\log
\epsilon$(Li)$=1.5$. This is clearly lower than the values found in
metal-poor main sequence or subgiant stars with similar effective
temperatures ($T_{\rm eff}\sim 6200$~K). For instance,
\citet{Ryanetal:1999} determined Li abundances of $\log
\epsilon$(Li)$=2.0$ to 2.2 for 22 main-sequence turnoff stars with
$-3.6<$[Fe/H]$<-2.3$ and 6050~K$< T_{\rm eff} < 6350$~K. The
dependence of the Li abundance on metallicity found by
\citet{Ryanetal:1999} is represented as $\log \epsilon$ (Li)$ = (2.447
\pm 0.066) + (0.118 \pm 0.023)$[Fe/H]. A simple extrapolation to
[Fe/H]$=-5.6$ (the LTE value) results in $\log \epsilon$ (Li)$=1.79$,
although there is at present no physical reason for such an
extrapolation.

The Li abundance of the previously most metal-poor dwarf CS~22876--032
is $\log \epsilon$(Li) $=2.03$ \citep{Norrisetal:2000}.  Even if
{\FAC} is a subgiant, no significant depletion of Li is expected,
given the result that a cooler metal-poor subgiant, HD~140283 ($T_{\rm
eff}\sim 5750$~K), has $\log \epsilon$(Li)$=2.1$
\citep[e.g.,][]{ford02}.

The Li abundances measured for these very metal-poor dwarfs are lower
than the primordial Li abundance expected from standard Big Bang
Nucleosynthesis models if one adopts the baryon density estimated from
the Wilkinson Microwave Anisotropy Probe \citep[WMAP;
][]{spergel03}. The explanations for this discrepancy are still
controversial \citep[e.g. ][]{coc04}, and no clear solution has yet
been obtained. This problem is beyond the scope of the present work,
although it clearly remains important to pursue.  The question to be
addressed here is the discrepancy between the Li abundance of {\FAC}
and other extremely metal-poor stars with similar effective
temperatures.

``Li-depleted" stars are known to exist among metal-poor,
main-sequence stars, although the number fraction of such stars is
quite small: about 5\%. \cite{Ryanetal:2002} investigated in detail
four ``Li-depleted" halo stars, which are main-sequence turnoff stars
that are extremely deficient in Li. They found three of them to
exhibit substantial line broadening, which they attributed to stellar
rotation ($v \sin i=5.5 - 8.3$~km~s$^{-1}$). They hypothesized that
the high rotational velocity is caused by mass and angular momentum
transfer across a binary system from an initially more massive donor,
which has also affected the surface Li abundance of the star currently
being observed.

The line widths of {\FAC}, and of {\GG}, for comparison purposes, were
reported in \S~\ref{sec:linewidth}. The FWHM of {\FAC} is 9.2~km~s$^{-1}$, on
average, for equivalent widths of 20--30~m{\AA}. The FWHM of lines of similar
strength in Li-normal stars studied by \citet{Ryanetal:2002} is 9~km~s$^{-1}$
(Figure 1 of their paper). The spectral resolution of our spectra
($R=60,000$) is slightly higher than those of \citet{Ryanetal:2002}
($R=43,000$ or 50,000). Given the effects of the difference in resolution,
which are 0.8--1.8~km~s$^{-1}$ for line widths of 9.2~km~s$^{-1}$, the
line broadening of {\FAC} might be slightly larger than those of ``Li-normal"
stars. This is, however, within the scatter of the line widths of
``Li-normal" stars. Indeed, the FWHM of the Mg lines measured for the
``Li-normal" star {\GG} (9.8~km~s$^{-1}$ at equivalent widths of
78--93~m{\AA}, see Table~\ref{tab:lw}) is also slightly larger than those in
``Li-normal" stars studied by \citet{Ryanetal:2002}, if the difference of
spectral resolution is taken into account. 
It should be noted that this result relies on the measurements for only two
Mg lines. In order to derive a definitive conclusion, estimates from higher
quality spectra and/or a larger number of lines are strongly desired. \\

Our conclusion here is that no clear excess broadening by rotation or
macroturbulence, with respect to ``Li-normal" stars, is found in
{\FAC}, and the cause for the lower Li abundance of this star than
those of other metal-poor main-sequence stars or warm subgiants has
not yet been identified. It has to be remarked, however, that none of
the other metal-poor main-sequence stars has such large overabundances
of C and N, so it is far from obvious that we should expect {\FAC} to
have the same abundance of such a fragile element like Li.

\subsection{Neutron-capture elements}\label{sec:sr}

Among elements heavier than the iron group, only Sr is detected in
{\FAC}. Singly-ionized Sr has strong resonance doublet lines in the blue
region, which enable one to measure its abundance in very metal-poor
stars. While the Sr/Fe ratio of {\FAC} is remarkably high ([Sr/Fe] $\sim +1$),
given its low Fe abundance, the Sr abundance ($\log \epsilon$(Sr)$ \sim -1.7$)
itself falls within the large dispersion of Sr abundances ($\log
\epsilon$(Sr)$ = -3$ to 0) found in very metal-poor stars (e.g., Figure 6 of
Aoki et al. 2005).  We note that the upper limit of Sr abundance for {\CBB}
($\log \epsilon$(Sr)$ <-2.83$, [Sr/Fe]$ < -0.52$) is significantly lower than
the value found here for {\FAC}. The large scatter of Sr/Fe found in very
metal-poor stars suggests that the origin of this neutron-capture element is
quite different from that of Fe in the early Galaxy. The excess of Sr in the
Fe-deficient star {\FAC} supports this suggestion.

Singly-ionized Ba also has strong resonance lines, at 4554~{\AA} and
4934~{\AA}, which are not seen in the spectrum of {\FAC}. The upper
limit on the Ba abundance estimated from the \ion{Ba}{2} 4554~{\AA}
line is [Ba/Fe]$\lesssim +1.5$. This limit is weak, given that most
extremely metal-poor stars with [Fe/H]$<-3$ have [Ba/Fe]$<0$
(e.g. McWilliam 1998; Honda et al. 2004). The limit for the
Sr/Ba ratio ([Sr/Ba]$> -0.4$), however, provides an important
constraint on the origin of neutron-capture elements in this
object. The value of Sr/Ba produced by the main s-process at low
metallicity is known to be very low ([Sr/Ba]$\lesssim -1$)
\citep[e.g.][]{Aokietal:2002c}. This is interpreted as the result of a
deficiency of seed nuclei with respect to neutrons that are expected
to be provided from $^{13}$C($\alpha,n$)$^{16}$O, independent of
metallicity. Hence, the main s-process is unlikely to have been
responsible for the Sr in {\FAC}. The weak s-process, which was
introduced to explain the light s-process nuclei ($A\lesssim 90$) in
solar-system material, produces Sr with essentially no Ba. Massive,
core He-burning stars are regarded as the astrophysical sites of this
process. Theoretical calculations, however, suggest that it is
inefficient at low metallicity, because the neutron source expected
for the process, $^{22}$Ne($\alpha,n$)$^{25}$Mg, has a strong
metallicity dependence \citep[e.g. ][]{prantzos90}.

The Sr/Ba ratio produced by the r-process is best estimated from the
value observed in the so-called r-II stars, stars with very large
enhancements of r-process nuclei with respect to other metals:
[Sr/Ba]$=-0.41$ in CS~22892--052 \citep{sneden03}, $-0.52$ in
CS~31082--001 \citep{hill02}, and $-0.46$ in CS~29497--004
\citep{HERESpaperI}. Although the Sr/Ba ratios in other metal-poor
stars exhibit a large scatter, the concordance of the ratio in these
three stars suggests that the value produced by the r-process is
[Sr/Ba]$=-0.5$ to $-0.4$. This value is not inconsistent with the
lower limit of [Sr/Ba] in {\FAC}. Therefore, it is at least feasible
that the Sr in {\FAC} might have originated from the r-process. Type
II supernovae are the most promising sites for this process, although
other possibilities have also been proposed \citep[][ and references
therein]{truran02}. For the purpose of further discussion we refer to
the r-process that yields heavy neutron-capture elements as the ``main
r-process''.

Another possible process has recently been proposed to account for the
production of light neutron-capture elements in the early
Galaxy. There exists a number of metal-poor stars with high Sr/Ba
abundance ratios that cannot be explained by the r-process mentioned
above (McWilliam 1998; Honda et al. 2004). \citet{aoki05} showed that
such stars are particularly evident at extremely low metallicity
([Fe/H]$\lesssim -3$). They also demonstrated that the weak s-process
cannot explain the abundance ratio of Y/Zr of these objects. To
explain these stars, a process that has provided light neutron-capture
elements, including Sr, with little (or no) heavy neutron-capture
elements, is required. Although the site and mechanism of this process
are still unknown, its characteristics have been studied by
\citet{truran02}, \citet{travaglio04}, and \citet{aoki05}. In order to
identify the origin of Sr in {\FAC}, a stronger constraint on the Ba
abundance is necessary. If the Ba abundance is found to be half that
of the upper limit determined by the present work, the main r-process
will be excluded as the source of Sr in {\FAC}. Further information on
other neutron-capture elements would also be useful. The spectral
features of elements other than Sr and Ba, however, are expected to be
very weak, and extremely high-quality data will be required in order
to detect them.

We note that the Ba abundance of {\GG} ([Ba/Fe]$=-0.25$) is significantly
higher than that of stars with similar extremely low metallicity ([Fe/H]$\sim -3.2$, Aoki
et al. 2005). Moreover, the Sr/Ba ratio of this object ([Sr/Ba]$=+0.38$) is
clearly higher than expected from the main r-process ($\sim -0.4$). These
facts suggest a unique history of enrichment of neutron-capture elements in
this object.

\section{Interpretations of the chemical abundance patterns in HMP stars} 

{\FAC} shares important chemical abundance characteristics with
{\CBB}, (extreme deficiency of iron-group elements and large
overabundance of carbon), coupled with significant
differences. Several scenarios have been proposed to explain the
abundance pattern of {\CBB} following its discovery by
\citet{HE0107_Nature}. Perhaps the most interesting question is
whether this object was an initially metal-free, low-mass population
III star, or an extreme population II one. From this point of view,
the scenarios proposed for {\CBB} might be classified into two groups.
One is to assume that it was initially metal-free, and the accretion
of a small amount of metals from the ISM provided the iron-group
elements currently observed at its surface \citep{yoshii81}. In this case, an
additional source is required to explain its high abundances of C, N,
and O \citep{HE0107_Nature,Besselletal:2004}. The other is to assume
supernova(e), whose progenitor(s) might be first-generation massive
stars, which have provided small amounts of iron-group elements
together with large quantities of C, N, and O.

The discovery of {\FAC} provides new constraints on these models, if
we assume that the abundance patterns of these two stars have a common
origin.  The challenge is that the important differences between the
two stars have to be explained self-consistently. The differences are
(1) their evolutionary status, (2) the abundance ratios of N, Na, Mg
and Al with respect to Fe, and (3) the abundance ratio of Sr. The
non-detection of Li in {\FAC} is another constraint that is not
available from {\CBB} because of its evolutionary status.

\subsection{Population III scenarios}\label{sec:pop3}

We first discuss the possibility that these objects were formed from a
metal-free cloud of gas. In this case, the large overabundances of C
and N need to be explained by sources other than accretion from the
ISM. Self-enrichment of these elements in {\CBB} was discussed as one
possibility \citep[e.g.][]{Shigeyamaetal:2003}. The evolutionary
status of {\FAC} (subgiant or main sequence), however, clearly
excludes this hypothesis. Self-enrichment of C and N is also unlikely
in {\CBB} \citep{weiss04,picardi04}, because of its high C/N and
$^{12}$C/$^{13}$C ratios \citep{HE0107_ApJ}.

A remaining possibility is mass transfer in a binary system where the
primary star provides the secondary with C and N when the former was
in its AGB phase of evolution. After the mass transfer, the primary
star evolved to become a faint white dwarf, and only the secondary is
currently observable. \citet{suda04} discussed this possibility in
considerable detail.  According to their calculations of
nucleosynthesis in AGB stars, the excess of O and Na found in {\CBB}
is the result of $^{13}$C($\alpha,n$)$^{16}$O and neutron-capture
starting from $^{20}$Ne, following the
$^{16}$O$(n,\gamma)^{17}$O$(\alpha,n)$ reactions.  The large
overabundances of Mg and Al in {\FAC} also need to be explained as the
contribution from AGB stars. \citet{suda04} demonstrated possible
production of these elements by neutron-capture processes such as
those involving Na. Further calculations are needed to investigate
whether such processes can explain the abundance pattern of elements
from C to Al in {\FAC}, especially its C/N ratio. It should also be
noted that these neutron-capture processes primarily yield heavy Mg
isotopes ($^{25}$Mg and $^{26}$Mg), while $^{24}$Mg is produced by
type II supernovae. Unfortunately, the isotope fractions of Mg are not
measurable for these objects, since no MgH feature has yet been
detected.

The scenario of mass transfer from an AGB star has the potential to
explain the non-detection of Li in {\FAC}, because it is expected to
be depleted in most of evolved stars (see, however, \citet{iwamoto04}
for possible Li enhancement in the N-enhanced AGB stars). Indeed, this
is the explanation for Li-depleted main-sequence stars proposed by
\citet{Ryanetal:2002}, although no signature of rapid rotation or
binarity is found in {\FAC}. On the other hand, the high Sr abundance
cannot be explained as a result of AGB nucleosynthesis. As seen in
\S~\ref{sec:sr}, the s-process expected in metal-poor AGB stars
produces low Sr/Ba ratios, which are inconsistent with the lower limit
to the Sr/Ba ratio in {\FAC}. Another explanation is therefore
required for the observed Sr abundance in this object.

Since the evolutionary status of {\FAC} and {\CBB} is quite different,
the depths of their surface convection zones should be significantly
different (presumably by more than two orders of magnitude).
Therefore, if their surfaces had been similarly contaminated by the
ISM, the Fe abundance of {\CBB} would be expected to be much lower
than that of {\FAC}. Their comparable Fe abundances disagree with this
expectation.  The problem, however, might perhaps be understood if
mass transfer in a binary system is assumed, in which the surfaces of
{\CBB} and {\FAC} preserve material transferred from AGB companions.
Here, the Fe abundance ratio is primarily determined by the dilution
of Fe accreted from the ISM in AGB stars rather than in the stars we
are currently observing. This requires that the accretion from the ISM
would be significant enough to pollute the whole envelope of the AGB
donor. Accretion from the ISM was also investigated by \citet{suda04},
who estimated the effect inside the star forming region to be
significant.

Some recent theoretical studies \citep[][ and references
therein]{bromm04} on early star formation predict the formation of
exclusively massive and/or super-massive stars from primordial
clouds. Others suggest the possibility of low-mass star formation
\citep[e.g. ][]{Nakamura/Umemura:2001}. The confirmation of {\FAC} and
{\CBB} as low-mass population III objects would clearly have a great
impact on this discussion.

\subsection{Population II scenarios}\label{sec:pop2}

We now discuss the possibility that {\FAC} and {\CBB} were born from
material polluted by first-generation supernovae. If this scenario
pertains, then these objects should be considered extreme examples of
population II stars. Their abundance patterns, however, are radically
different from those found in most stars with [Fe/H]$>-4$, and
therefore require ``special" models of progenitor supernovae.

\citet{Umeda/Nomoto:2003} proposed a supernova model involving two
additional parameters -- mixing after explosive nucleosynthesis and
subsequent fallback onto the presumed collapsed remnant. Such a mixing
was actually observed in SN 1987A and found to take place in the
simulation of Rayleigh-Taylor instabilities during the explosion
\citep[see ][ for a review]{nomoto94}. Unusually large fallback can
result in small yields of metals other than C, N, and O. Mixing,
postulated to occur before the fallback, transfers a small amount of
iron-peak elements into the upper layers, which are ejected by the
explosion. By tuning these two parameters, their model for a
zero-metal 25M$_{\odot}$ progenitor successfully explains the large
overabundances of light elements and the deficiency of heavy elements
found in {\CBB}. Since the small yield of Fe indicates that only a
small amount of $^{56}$Ni is synthesized and thus the radioactive
energy input from the decay of $^{56}$Ni into $^{56}$Co and $^{56}$Fe
to power the supernova light curve tail is small.  Thus such models
are sometimes referred to as ``faint supernovae'' \citep{nomoto03}.
\citet{Umeda/Nomoto:2003} suggested that the two parameters (mixing
and fallback) might represent the phenomena occurring in jet-induced
aspherical supernova explosions \citep[see also][]{umeda05}.

One advantage of this model is that the observed differences in the
Mg/Fe ratios between {\FAC} and {\CBB} are easily explained by very
small variations of these parameters. Indeed, \citet{iwamoto05} quite
recently showed that a small difference in the explosion energy 
causes a difference in the fallback mass, 
which results in large differences of
abundance ratios of Na, Mg, and Al with respect to Fe, and succeeded
to explain the abundance pattern of {\FAC}. This idea had also been
already applied by \citet{Umeda/Nomoto:2003} to the carbon-enhanced
stars CS~22949--037 and CS~29498--043, which exhibit overabundances of
$\alpha$-elements along with C, N, and O
\citep{Norrisetal:2001,Depagneetal:2002,aoki04}. The discovery of
{\FAC} supports the connection between {\CBB} and these
carbon-enhanced stars. The excess of odd-numbered elements is not
simply explained by supernova models. However, \citet{iwamoto05}
showed that these can be explained by including the effect of
overshooting in the convective carbon-burning shell in the progenitor
models.
  
Another advantage of the model is that such faint supernovae provide
large amounts of light elements, which could be efficient cooling
sources in the formation of second-generation low-mass stars
\citep{Umeda/Nomoto:2003}. It should be appreciated that {\FAC} and
{\CBB}, their deficiency of iron-group elements notwithstanding, have
relatively high ``total metallicities'', because of their carbon
excesses. \citet{Bromm/Loeb:2003} showed that the carbon abundance of
{\CBB} is sufficient to initiate low-mass star formation.

The excesses of C, N, and O are also expected in the yields from the
rotating, massive stars. \citet{meynet05} quite recently reported the
effect of rotation on their model calculations for 60~$M_{\odot}$ stars
with [Fe/H]$=-6.1$. They predict heavy mass loss for the rotating
stars, and the large enhancements of C, N, and O in the
winds. Moderate excesses of Na and Al are also found. The excesses of
these elements, in particular of N, Na and Al, at least qualitatively
explain the abundance characteristics found in {\FAC} and {\CBB}.

A critical unexplained problem is the excess of Sr found in
{\FAC}. Production of neutron-capture elements is not included in the
\citet{Umeda/Nomoto:2003} models, or other supernova models. One
possibility might be the r-process occurring in accretion disks or
jets expected in anisotropic supernovae models including rotation
\citep[e.g. ][]{fujimoto04,surman05}. Another possibility is a
so-called ``weak'' r-process, which might occur in a neutron star wind
before the neutron star collapses into a black hole due to fallback
\citep{wanajo05}. A second no less enigmatic
difficulty is the Li depletion in {\FAC}, for which no clear
understanding is provided by the population II scenario. That said, an
{\it ad hoc} explanation might be provided by internal rotation
effects not reflected in our spectra \citep[e.g. ][]{pinsonneault99},
because of aspect effects.

\subsection{Further constraints expected from future observations}

Our tentative conclusion is that the abundance patterns of {\FAC} and
{\CBB} can be explained by either the ``faint supernova" model, or by
binary mass transfer along with accretion of metals from the ISM. In
order to obtain further constraints on these models, the following
observational data for {\FAC} would be very useful:

{\it Str\"{o}mgren photometry:} In order to constrain the evolutionary
status, the Str\"{o}mgren uvby-{$\beta$} photometry would be very
useful \citep[e.g. ][]{nissen89}.

{\it Fe lines:} Since the number of detected \ion{Fe}{1} lines is
still small, a search for more lines, in particular those in the near-UV
range, would provide a more accurate measurement of the Fe
abundance. Though detections of \ion{Fe}{2} lines will be very
difficult, because of their expected weakness, a stronger upper limit for these
lines would be useful to provide further constraints on gravity from
the ionization balance technique.

{\it O abundance:} Given the abundance pattern of {\FAC}, a large
excess of O is expected. The abundance measurement for this element is
key to constraining the nucleosynthesis models of supernovae and AGB
stars. Further observations of OH lines in the near-UV region and of
the near infrared \ion{O}{1} triplet will provide a stronger
constraint on the [O/Fe], as discussed in \S \ref{sec:on}.

{\it Ba abundance:} The abundance ratio of Sr/Ba, or its lower limit, is
essential for an understanding of the origin of the overabundance of Sr in
this object, and to discriminate between proposed nucleosynthesis models
(see \S \ref{sec:sr}). Further observations of the \ion{Ba}{2} resonance lines
at 4554~{\AA} and 4934~{\AA} will yield a stronger constraint on the Ba
abundance.

{\it Li abundance:} Although the position of Li below the Spite plateau value
in this object is already clear, determination of its abundance, or at least
a stronger upper limit, will be useful for understanding the reason for
its apparent depletion\footnote {While not emphasized in the
present work, Li abundances for additional dwarfs having [Fe/H] as low as that
of {\FAC} are needed to examine the possible cosmological implications of the
present result.}.

{\it Measurements of line widths:} In order to constrain the reason for the
Li-depletion, accurate measurements of line widths and shapes are important
(see \S~\ref{sec:lidep}). For this purpose, measurements for a larger number of
lines than that used in the present work, from high S/N ratio spectra, are
required.

{\it Radial velocity monitoring:} This is a key for examining scenarios
involving mass transfer across a binary system. \citet{suda04} estimated that
a long orbital period is sufficient to explain the mass transfer required for
the abundance patterns of light elements in {\CBB}. A long term monitoring
program with high accuracy will be necessary to examine this model for
{\FAC}.

\section{Implications for future surveys for ultra metal-poor stars}

The discovery of {\FAC} and {\CBB} provide important lessons for
future searches for HMP stars with $\mbox{[Fe/H]}<-5.0$. Given that
the search for such objects is (presently) based on the detection of
the \ion{Ca}{2}~K line, it is worthwhile to examine the complications
that exist for the application of this technique to stars at such
extreme abundances. In the moderate-resolution follow-up spectrum of
{\CBB}, features at the position of the \ion{Ca}{2}~K line yielded
only an upper limit for its iron abundance of $\mbox{[Fe/H]}<-4.0$.
High-resolution spectra obtained later revealed that, at
medium-resolution (FWHM $\sim$ 2\,{\AA}, $R\sim 2,000$), the
\ion{Ca}{2}~K line is blended with a CH line, which would have
resulted in an erroneous identification of the \ion{Ca}{2}~K line and
a considerable overestimate of the star's iron abundance, had it been
cooler and/or its C abundance higher.

In the case of {\FAC}, contamination of the \ion{Ca}{2}~K line by
interstellar absorption, which cannot be resolved from the intrinsic
\ion{Ca}{2}~K line at moderate spectral resolution, and the Ca
overabundance of about $\mbox{[Ca/Fe]}=+0.8$\,dex (from the
\ion{Ca}{2} K line) led to an overestimate of [Fe/H] by as much as
1.5\,dex. The Ca overabundance is relevant here, because in the
calibration of the \ion{Ca}{2}~K index \texttt{KP} of
\citet{Beersetal:1999}, it is implicitly assumed that
$\mbox{[Ca/Fe]}=+0.4$ for $\mbox{[Fe/H]} < -1.5$. While this
assumption is justified for the vast majority of metal-poor stars
found in surveys such as the HES, it can lead to overestimates of
[Fe/H] for the lowest metallicity stars, because they may have larger
overabundances of Ca. Furthermore, the rarity of stars at
$\mbox{[Fe/H]}<-3.5$ results in a severe shortage of calibration
objects at the lowest abundances, resulting in potentially large
uncertainties in the \texttt{KP}-based [Fe/H] measurement in this
metallicity range.

For these reasons, spectra with higher resolution (e.g. $R\sim
20,000$) are needed for reliable [Fe/H] estimates of objects having
$\mbox{[Fe/H]}\lesssim -3.5$, in order to identify stars among them
with $\mbox{[Fe/H]}<-5$.

\section{Summary and concluding remarks}

A comprehensive abundance analysis has been carried out for {\FAC},
the most iron-deficient star known, using the high resolution spectrum
obtained with Subaru/HDS. Our analyses revealed that (1) the NLTE
corrected iron abundance is [Fe/H] $=-5.45$, slightly lower than that
of {\CBB}, and more than 1~dex lower than those of all other
metal-poor stars; (2) carbon shows a significant overabundance
([C/Fe]$\sim +4.0$), similar to {\CBB}; (3) light elements (Na, Mg and
Al) show moderate enhancements, while N has a remarkable overabundance
([N/Fe]$\gtrsim +4.0$); (4) the light neutron-capture element Sr also
shows an enhancement ([Sr/Fe]$\sim 1.0$); (5) the upper limit of Li
abundance ($\log \epsilon$(Li)$<1.5$) is below the value of Spite
plateau. No significant change in radial velocity has been found in
our monitoring in the past year.

The combination of extremely high carbon abundance with outstandingly
low iron abundance in {\FAC} and {\CBB} clearly distinguishes these
two objects from other metal-poor stars. Several models to explain
such carbon-enhanced, iron-deficient stars have been proposed. The
important differences of abundance patterns from N to Al, as well as
of Sr abundances, between {\FAC} and {\CBB} provide new constraints on
models of nucleosynthesis processes in the first generation objects
that were responsible for metal enrichment at the earliest times.  

In order to give stronger constraints on these models, further
abundance studies for {\FAC} would be very useful. In particular,
determination of Li, O, and Ba abundances, or  stronger upper
limits, are urgently needed. To address these issues, a new high
resolution spectrum has been taken with VLT/UVES (Frebel et al., in
preparation). Further monitoring of radial velocities of {\FAC}, as
well as of {\CBB}, to investigate their binarity, will also provide
important information for the understanding of origins of the peculiar
abundance patterns of these objects.

In order to obtain a comprehensive picture of nucleosynthesis and star
formation in the early Universe, extensive abundance studies for a
larger sample of ultra metal-poor stars are required. In particular,
the apparent metallicity gap between [Fe/H] $= -4$ and $-5$ found in the
present sample should be confirmed.

\acknowledgments

We are grateful to Arto J\"arvinen, Jyri N\"ar\"anen, and Brian Krog
for obtaining additional photometry for {\FAC} with the 0.9\,m NOT
telescope.  We thank Akito Tajitsu, the support astronomer of the HDS,
for special support for our Subaru observations. A.F. thanks the
National Astronomical Observatory of Japan for its hospitality. This
research made extensive use of the Vienna Atomic Line Database (VALD),
and the Abstract Service of NASA's Astrophysics Data System.  W.A.,
K.N., M.Y.F., and Y.Y. are supported by a Grant-in-Aid for Science
Research from JSPS (grant 152040109). A.F., J.E.N., and M.A. are
supported by the Australian Research Council through grant DP0342613,
while A.F. acknowledges travel funds awarded by the Astronomical
Society of Australia. N.C. acknowledges financial support from
Deutsche Forschungsgemeinschaft under grants Ch~214/3-1 and
Re~353/44-2. T.C.B.  acknowledges partial support for this work from
grants AST 00-98508, AST 00-98549, AST 04-06784, and PHY 02-16783,
Physics Frontier Centers/JINA: Joint Institute for Nuclear
Astrophysics, awarded by the US National Science Foundation. A.F.,
N.C., and J.E.N. express gratitude to JINA for sponsorship of their
visits to Michigan State University, during which useful discussions
of this work took place.  P.B. and K.E. acknowledge support from the
Swedish Reseach Council.


\begin{thebibliography}{106}
\expandafter\ifx\csname natexlab\endcsname\relax\def\natexlab#1{#1}\fi

\bibitem[{Akerman {et~al.}(2004)Akerman, Carigi, Nissen, Pettini, \&
  Asplund}]{akerman04}
Akerman, C.~J., Carigi, L., Nissen, P.~E., Pettini, M., \& Asplund, M. 2004,
  A\&A, 414, 931

\bibitem[{Ali \& Griem(1966)}]{Ali/Griem:1966}
Ali, A., \& Griem, H. 1966, Phys. Rev., 144, 366

\bibitem[{Alonso {et~al.}(1996)Alonso, Arribas, \&
  {Mart{\'\i}nez-Roger}}]{Alonsoetal:1996}
Alonso, A., Arribas, S., \& {Mart{\'\i}nez-Roger}, C. 1996, A\&A, 313, 873

\bibitem[{Aoki {et~al.}(2005)Aoki, Honda, Beers, Kajino, Ando, Norris,
  Ryan, Izumiura, Sadakane, \& Takada-Hidai}]{aoki05} Aoki, W., Honda,
  S., Beers, T. C., Kajino, T., Ando, H., Norris, J. E., Ryan, S. G.,
  Izumiura, H., Sadakane, K., \& Takada-Hidai, M. 2005, ApJ, in press
  (astro-ph/0503032)

\bibitem[{Aoki {et~al.}(2002{\natexlab{a}})Aoki, Norris, Ryan, Beers,
  \& Ando}]{Aokietal:2002d} Aoki, W., Norris, J. E., Ryan, S. G.,
  Beers, T. C., \& Ando, H. 2002{\natexlab{a}}, ApJ, 576, L141

\bibitem[{Aoki {et~al.}(2004)Aoki, Norris, Ryan, Beers, Christlieb,
  Tsangarides, \& Ando}]{aoki04} Aoki, W., Norris, J. E., Ryan, S. G.,
  Beers, T. C., Christlieb, N., Tsangarides, S., \& Ando, H. 2004,
  ApJ, 608, 971

\bibitem[{Aoki {et~al.}(2002{\natexlab{b}})Aoki, Ryan, Norris, Beers,
  Ando, \& Tsangarides}]{Aokietal:2002c} Aoki, W., Ryan, S. G.,
  Norris, J. E., Beers, T. C., Ando, H., \& Tsangarides, S.
  2002{\natexlab{b}}, ApJ, 580, 1149

\bibitem[{{Asplund} {et~al.}(2005b){Asplund}, {Grevesse}, {Sauval}, {Allende
  Prieto}, \& {Blomme}}]{asplund05c}
{Asplund}, M., {Grevesse}, N., {Sauval}, A.~J., {Allende Prieto}, C., \&
  {Blomme}, R. 2005b, \aap, 431, 693

\bibitem[{Asplund {et~al.}(2005a)Asplund, Grevesse, \& Sauval}]{asplund05}
Asplund, M., Grevesse, N., \& Sauval, A. J. 2005a, ASP Conf. Ser. 336:
Cosmic Abundances as Records of Stellar Evolution and
Nucleosynthesis, Ed, T. G. Barnes \& F. N. Bash, 25 

\bibitem[Asplund et al. (1999)]{asplund99} Asplund, M., Nordlund, {\AA}.,
Trampedach, R., \& Stein, R. F. 1999, \aap, 346, L17

\bibitem[Asplund (2004)]{asplund04} Asplund, M. 2004, Mem. S. A. It.,
75, 300

\bibitem[{Barklem {et~al.}(2000)Barklem, Piskunov, \&
  O'Mara}]{Barklemetal:2000}
Barklem, P. S., Piskunov, N., \& O'Mara, B. J. 2000, A\&A, 363, 1091

\bibitem[{Barklem {et~al.}(2002)Barklem, Stempels, {Allende Prieto},
  Kochukhov, Piskunov, \& O'Mara}]{Barklemetal:2002} Barklem, P. S.,
  Stempels, H. C., {Allende Prieto}, C., Kochukhov, O. P., Piskunov,
  N., \& O'Mara, B. J. 2002, A\&A, 385, 951

\bibitem[Baum\"{u}ller \& Gehren (1997)]{baumuller97} Baum\"{u}ller,
D., \& Gehren, T. 1997, \aap, 325, 1088

\bibitem[{Beers(1999)}]{TimTSS} Beers, T.~C. 1999, in ASP Conf. Ser.,
Vol. 165, The Third Stromlo Symposium: The Galactic Halo,
ed. B.~Gibson, T.~Axelrod, \& M.~Putman, 202

\bibitem[{Beers \& Christlieb(2005)}]{beers05}
Beers, T.~C., \& Christlieb, N. 2005, ARA\&A, in press

\bibitem[{Beers {et~al.}(1985)Beers, Preston, \& Shectman}]{BPSI}
Beers, T.~C., Preston, G.~W., \& Shectman, S.~A. 1985, AJ, 90, 2089

\bibitem[{Beers {et~al.}(1992)Beers, Preston, \& Shectman}]{BPSII}
---. 1992, AJ, 103, 1987

\bibitem[{Beers {et~al.}(1999)Beers, Rossi, Norris, Ryan, \&
  Shefler}]{Beersetal:1999}
Beers, T.~C., Rossi, S., Norris, J.~E., Ryan, S.~G., \& Shefler, T. 1999, AJ,
  117, 981

\bibitem[{Bessell(1983)}]{Bessell:1983}
Bessell, M.~S. 1983, PASP, 95, 480

\bibitem[{Bessell {et~al.}(2004)Bessell, Christlieb, \&
  Gustafsson}]{Besselletal:2004}
Bessell, M., Christlieb, N., \& Gustafsson, B. 2004, ApJ, 612, L61

\bibitem[{Blackwell {et~al.}(1989)Blackwell, Booth, D.Petford, \&
  Laming}]{blackwell89} Blackwell, D.~E., Booth, A. J., Petford,
  A. D., \& Laming, J.~M. 1989, MNRAS, 236, 235

\bibitem[{{Bromm} \& {Larson}(2004)}]{bromm04}
{Bromm}, V., \& {Larson}, R.~B. 2004, \araa, 42, 79

\bibitem[{Bromm \& Loeb(2003)}]{Bromm/Loeb:2003}
Bromm, V., \& Loeb, A. 2003, Nature, 425, 812

\bibitem[{Brown(1987)}]{brown87}
Brown, J.~A. 1987, ApJ, 317, 701

\bibitem[{Burstein \& Heiles(1982)}]{Burstein/Heiles:1982}
Burstein, D., \& Heiles, C. 1982, AJ, 87, 1165

\bibitem[{{Carswell} {et~al.}(1987){Carswell}, {Webb}, {Baldwin}, \&
  {Atwood}}]{VPFIT}
{Carswell}, R.~F., {Webb}, J.~K., {Baldwin}, J.~A., \& {Atwood}, B. 1987, \apj,
  319, 709

\bibitem[{{Castelli} {et~al.}(1997){Castelli}, {Gratton}, \&
  {Kurucz}}]{castelli97}
{Castelli}, F., {Gratton}, R.~G., \& {Kurucz}, R.~L. 1997, \aap, 318, 841

\bibitem[{Cayrel {et~al.}(2004)Cayrel, Depagne, Spite, Hill, Spite,
  Francois, Beers, Primas, Andersen, Barbuy, Bonifacio, Molaro, \&
  Nordstr\"om}]{Cayreletal:2004} Cayrel, R., et al. 2004,
  A\&A, 416, 1117

\bibitem[{Christlieb(2003)}]{Christlieb:2003}
Christlieb, N. 2003, Rev. Mod. Astron., 16, 191, astro-ph/0308016

\bibitem[{{Christlieb} {et~al.}(2004a){Christlieb}, {Beers}, {Barklem},
  {Bessell}, {Hill}, {Holmberg}, {Korn}, {Marsteller}, {Mashonkina}, {Qian},
  {Rossi}, {Wasserburg}, {Zickgraf}, {Kratz}, {Nordstr{\" o}m}, {Pfeiffer},
  {Rhee}, \& {Ryan}}]{HERESpaperI}
{Christlieb}, N. et al. 2004a,  \aap, 428, 1027 

\bibitem[{Christlieb {et~al.}(2002)Christlieb, Bessell, Beers,
  Gustafsson, Korn, Barklem, Karlsson, Mizuno-Wiedner, \&
  Rossi}]{HE0107_Nature} Christlieb, N., Bessell, M. S., Beers, T. C.,
  Gustafsson, B., Korn, A. J., Barklem, P. S., Karlsson, T.,
  Mizuno-Wiedner, M., \& Rossi, S. 2002, Nature, 419, 904

\bibitem[{Christlieb {et~al.}(2004b)Christlieb, Gustafsson, Korn,
  Barklem, Beers, Bessell, Karlsson, \& Mizuno-Wiedner}]{HE0107_ApJ}
  Christlieb, N., Gustafsson, B., Korn, A. J., Barklem, P. S., Beers,
  T. C., Bessell, M. S., Karlsson, T., \& Mizuno-Wiedner, M. 2004b,
  ApJ, 603, 708

\bibitem[{Coc {et~al.}(2004)Coc, Vangioni-Flam, Descouvemont, Adahchour, \&
  Angulo}]{coc04}
Coc, A., Vangioni-Flam, E., Descouvemont, P., Adahchour, A., \& Angulo, C.
  2004, ApJ, 600, 544

\bibitem[{Cohen {et~al.}(2002)Cohen, Christlieb, Cohen, Gratton, \&
  Carretta}]{Cohenetal:2002} Cohen, J. G., Christlieb, N., Beers,
  T. C., Gratton, R., \& Carretta, E. 2002, AJ, 124, 470

\bibitem[{Cohen {et~al.}(2004)Cohen, Christlieb, McWilliam, Shectman,
  Thompson, Wasserburg, Ivans, an~dTorgny Karlsson, \&
  Melendez}]{Cohenetal:2004} Cohen, J. G., Christlieb, N., McWilliam,
  A., Shectman, S., Thompson, I., Wasserburg, G. J., Ivans, I., Dehn,
  M., Karlsson, T., \& Melendez, J. 2004, ApJ, 612, 1107

\bibitem[{Cutri {et~al.}(2003)Cutri, Skrutskie, {van Dyk}, Beichman,
  Carpenter, Chester, Cambresy, Evans, Fowler, Gizis, Howard, Huchra,
  Jarrett, Kopan, Kirkpatrick, Light, Marsh, McCallon, Schneider,
  Stiening, Sykes, Weinberg, Wheaton, Wheelock, \&
  Zacharias}]{Cutrieetal:2003} Cutri, R. et al. 2003, 2MASS All-Sky
  Catalog of Point Sources, Tech.  rep., Infrared Processing and
  Analysis Center, vizieR Online Data Catalog II/2246


\bibitem[{Depagne {et~al.}(2002)Depagne, Hill, Spite, Spite, Plez,
  Beers, Barbuy, Cayrel, Andersen, Bonifacio, Fran{\c c}ois,
  Nordstr{\" o}m, \& Primas}]{Depagneetal:2002} Depagne, E., Hill, V.,
  Spite, M., Spite, F., Plez, B., Beers, T. C., Barbuy, B., Cayrel,
  R., Andersen, J., Bonifacio, P., Fran{\c c}ois, P., Nordstr{\" o}m,
  B., \& Primas, F. 2002, A\&A, 390, 187

\bibitem[{Ecuvillon {et~al.}(2004)Ecuvillon, Israelian, Santos, Mayor,
  L\'{o}pez, \& Randich}]{ecuvillon04} Ecuvillon, A., Israelian, G.,
  Santos, N.~C., Mayor, M., Garc\'{i}a L\'{o}pez, R. J., \&
  Randich, S. 2004, A\&A, 418, 703

\bibitem[{Ford {et~al.}(2002)Ford, Jeffries, Smalley, Ryan, Aoki, Kawanomoto,
  James, \& Barnes}]{ford02}
Ford, A., Jeffries, R.~D., Smalley, B., Ryan, S.~G., Aoki, W., Kawanomoto, S.,
  James, D.~J., \& Barnes, J.~R. 2002, A\&A, 393, 617

\bibitem[{Fran\c{c}ois {et~al.}(2003)Fran\c{c}ois, Depagne, Hill,
  Spite, Spite, Plez, Beers, Barbuy, Cayrel, Andersen, Bonifacio,
  Molaro, Nordstr\"om, \& Primas}]{Francoisetal:2003} Fran\c{c}ois,
  P., Depagne, E., Hill, V., Spite, M., Spite, F., Plez, B., Beers,
  T., Barbuy, B., Cayrel, R., Andersen, J., Bonifacio, P., Molaro, P.,
  Nordstr\"om, B., \& Primas, F. 2003, A\&A, 403, 1105

\bibitem[{{Frebel} {et~al.}(2005){Frebel}, {Aoki}, {Christlieb},
  {Ando}, {Asplund}, {Barklem}, {Beers}, {Eriksson}, {Fechner},
  {Fujimoto}, {Honda}, {Kajino}, {Minezaki}, {Nomoto}, {Norris},
  {Ryan}, {Takada-Hidai}, {Tsangarides}, \& {Yoshii}}]{frebel05}
  {Frebel}, A. et al.  2005, \nat, 434, 871

\bibitem[{Fuhr {et~al.}(1988)Fuhr, Martin, \& Wiese}]{Fuhretal:1988}
Fuhr, J.~R., Martin, G.~A., \& Wiese, W.~L. 1988,
J. Phys. Chem. Ref. Data, 17, Suppl. 4


\bibitem[Fujimoto et al. (2004)]{fujimoto04} Fujimoto, S., 
Hashimoto, M., Arai, K., Matsuba, R. 2004, \apj, .614, 847


\bibitem[Gehren et al. (2004)]{gehren04} 
{Gehren}, T., {Liang}, Y.~C., {Shi}, J.~R., {Zhang}, H.~W., \& {Zhao},
G. 2004, \aap, 413, 1045

\bibitem[{Girard {et~al.}(2004)Girard, Dinescu, van Altena, Platais, Monet, \&
  L\'{o}pez}]{girard04}
Girard, T.~M., Dinescu, D.~I., van Altena, W.~F., Platais, I., Monet, D.~G., \&
  L\'{o}pez, C.~E. 2004, AJ, 127, 3060

\bibitem[{Gratton {et~al.}(1999)Gratton, Carretta, Eriksson, \&
  Gustafsson}]{Grattonetal:1999}
Gratton, R. G., Carretta, E., Eriksson, K., \& Gustafsson, G. 1999, A\&A, 350, 955

\bibitem[{{Grevesse} {et~al.}(1981){Grevesse}, {Biemont}, {Lowe}, \&
  {Hannaford}}]{grevesse81}
{Grevesse}, N., {Biemont}, E., {Lowe}, R.~M., \& {Hannaford}, P. 1981, Liege
  International Astrophysical Colloquia, 23, 211

\bibitem[{Grevesse {et~al.}(1996)Grevesse, Noels, \& Sauval}]{grevesse96}
Grevesse, N., Noels, A., \& Sauval, A.~J. 1996, ASP Conf. Ser., 99, Cosmic
  Abundances, ed. S. S. Holt \& G. Sonneborn (Cambridge Univ. Press), 117

\bibitem[{{Gustafsson} {et~al.}(1975){Gustafsson}, {Bell}, {Eriksson}, \&
  {Nordlund}}]{gustafsson75}
{Gustafsson}, B., {Bell}, R.~A., {Eriksson}, K., \& {Nordlund}, A. 1975, \aap,
  42, 407

\bibitem[{Heger \& Woosley(2002)}]{Heger/Woosley:2002}
Heger, A., \& Woosley, S. 2002, ApJ, 567, 532

\bibitem[{Hill {et~al.}(2002)Hill, Plez, Cayrel, Beers, Nordst\"{o}m,
  Andersen, Spite, Spite, Barbuy, Bonifacio, Depagne, Francois, \&
  Primas}]{hill02} Hill, V., et al. 2002, A\&A, 387, 560

\bibitem[{{Hobbs}(1974)}]{hobbs74}
{Hobbs}, L.~M. 1974, \apj, 191, 381

\bibitem[{Holweger \& M\"uller(1974)}]{Holweger/Mueller:1974}
Holweger, H., \& M\"uller, E. A. 1974, Solar Physics, 39, 19

\bibitem[{Honda {et~al.}(2004)Honda, Aoki, Kajino, Ando, Beers, Izumiura,
  Sadakane, \& Takada-Hidai}]{Hondaetal:2004b}
Honda, S., Aoki, W., Kajino, T., Ando, H., Beers, T. C., Izumiura, H., Sadakane,
  K., \& Takada-Hidai, M. 2004, ApJ, 607, 474

\bibitem[{Houdashelt {et~al.}(2000)Houdashelt, Bell, \&
  Sweigart}]{Houdasheltetal:2000} Houdashelt, M. L., Bell, R. A., \&
  Sweigart, A. V. 2000, AJ, 119, 1448

\bibitem[Howell et al. (2003)]{howell03} Howell, S. B., Everett,
M. E., Tonry, J. L., Pickles, A., Dain, C. 2003, PASP, 115, 1340

\bibitem[{Israelian {et~al.}(2004)Israelian, Ecuvillon, Rebolo,
  Garc\'{i}a-L\'{o}pez, Bonifacio, \& Molaro}]{israelian04} Israelian,
  G., Ecuvillon, A., Rebolo, R., Garc\'{i}a-L\'{o}pez, R., Bonifacio,
  P., \& Molaro, P. 2004, A\&A, 421, 649

\bibitem[Iwamoto {et~al.}(2004)]{iwamoto04} Iwamoto, N., Kajino, T.,
Mathews, G. J., Fujimoto, M. Y., \& Aoki, W. 2005, 602, 377

\bibitem[{Iwamoto {et~al.}(2005)Iwamoto, Umeda, Tominaga, Nomoto, \&
  Maeda}]{iwamoto05} Iwamoto, N., Umeda, H., Tominaga, N., Nomoto, K.,
  \& Maeda, K. 2005, Science, 309, 451

\bibitem[{K\c{e}pa {et~al.}(1996)K\c{e}pa, Para, Rytel, \& Zachwieja}]{kepa96}
K\c{e}pa, R., Para, A., Rytel, M., \& Zachwieja, M. 1996, J. Mol. Spectrosc.,
  178, 189

\bibitem[{Kerkhoff {et~al.}(1980)Kerkhoff, Schmidt, \& Zimmermann}]{kerkhoff80}
Kerkhoff, H., Schmidt, M., \& Zimmermann, P. 1980, Z. Phys. A., 298, 249

\bibitem[{Kim {et~al.}(2002)Kim, Demarque, Yi, \&
Alexander}]{Kimetal:2002} Kim, Y., Demarque, P., Yi, S., \& Alexander,
D. R. 2002, ApJS, 143, 499

\bibitem[{Korn \& Mashonkina(2005)}]{korn05}
Korn, A. J., \& Mashonkina, L. 2005, ASP Conf. Ser., IAU symposium
  228, From Lithium to Uranium: Elemental tracers of early cosmic evolution, in press

\bibitem[{Korn {et~al.}(2003)Korn, Shi, \& Gehren}]{Kornetal:2003}
Korn, A. J., Shi, J., \& Gehren, T. 2003, A\&A, 407, 691


\bibitem[{Kupka {et~al.}(1999)Kupka, Piskunov, Ryabchikova, Stempels, \&
  Weiss}]{vald99}
Kupka, F., Piskunov, N., Ryabchikova, T.~A., Stempels, H.~C., \& Weiss, W.~W.
  1999, A\&AS, 138, 119

\bibitem[{Kurucz(1993{\natexlab{a}})}]{kurucz93n15}
Kurucz, R.~L. 1993{\natexlab{a}}, Diatomic Molecular Data for Opacity
  Calculations. Kurucz CD-ROM No. 15. Cambridge, Mass (CD-ROM 15)

\bibitem[{Kurucz(1993{\natexlab{b}})}]{kurucz93}
Kurucz, R.~L. 1993{\natexlab{b}}, CD-ROM 13, ATLAS9 Stellar Atmospheres
  Programs and 2~km/s Grid (Cambridge: Smithsonian Astrophys. Obs.)

\bibitem[{Landolt(1992)}]{Landolt:1992}
Landolt, A. U. 1992, AJ, 104, 340

\bibitem[{Latham {et~al.}(2002)Latham, Stefanik, Torres, Davis, Mazeh, Carney,
  Laird, \& Morse}]{latham02}
Latham, D.~W., Stefanik, R.~P., Torres, G., Davis, R.~J., Mazeh, T., Carney,
  B.~W., Laird, J.~B., \& Morse, J.~A. 2002, AJ, 124, 1144

\bibitem[{Limongi {et~al.}(2003)Limongi, Chieffi, \&
  Bonifacio}]{Limongietal:2003}
Limongi, M., Chieffi, A., \& Bonifacio, P. 2003, ApJ, 594, L123

\bibitem[{McWilliam(1998)}]{Mcwilliam:1998}
McWilliam, A. 1998, AJ, 115, 1640

\bibitem[{McWilliam {et~al.}(1995)McWilliam, Preston, Sneden, \&
  Searle}]{McWilliametal:1995b}
McWilliam, A., Preston, G. W., Sneden, C., \& Searle, L. 1995, AJ, 109, 2757

\bibitem[{Meynet {et~al.}(2005)Meynet, Ekstr\"{o}m, \& Maeder}]{meynet05}
Meynet, G., Ekstr\"{o}m, S., \& Maeder, A. 2005, A\&A, in press

\bibitem[{Morton(1991)}]{morton91}
Morton, D.~C. 1991, ApJS, 77, 119

\bibitem[{Munari \& Zwitter(1997)}]{Munari/Zwitter:1997}
Munari, U., \& Zwitter, T. 1997, A\&A, 318, 269

\bibitem[{Nakamura \& Umemura(2001)}]{Nakamura/Umemura:2001}
Nakamura, F., \& Umemura, M. 2001, ApJ, 548, 19

\bibitem[{{Nissen} \& {Schuster}(1991)}]{nissen89}
{Nissen}, P.~E., \& {Schuster}, W.~J. 1991, \aap, 251, 457

\bibitem[{Noguchi {et~al.}(2002)Noguchi, Aoki, \& et~al.}]{noguchi02}
Noguchi, K., Aoki, W., Kawanomoto, S. et~al. 2002, PASJ, 54, 855

\bibitem[Nomoto et al. (1994)]{nomoto94} Nomoto, K., Shigeyama, T.,
Kumagai, S., Yamaoka, H., and Suzuki, T.  1994, in Supernovae, Les
Houches Session LIV, ed. S.A. Bludman, R. Mochkovitch, \&
J. Zinn-Justin (Amsterdam: North-Holland), 489

\bibitem[Nomoto et al. (2003)]{nomoto03} Nomoto, K., Maeda, K., Umeda,
H., Ohkubo, T., Deng, J., \& Mazzali, P. 2003, in IAU Symposium 212, A
Massive Star Odyssey, from Main Sequence to Supernova, eds. K.A. van
der Hucht, A. Herrero, \& C. Esteban (San Francisco: ASP), 395.

\bibitem[{Norris {et~al.}(2000)Norris, Beers, \& Ryan}]{Norrisetal:2000}
Norris, J.~E., Beers, T.~C., \& Ryan, S.~G. 2000, ApJ, 540, 456

\bibitem[{Norris {et~al.}(2001)Norris, Ryan, \& Beers}]{Norrisetal:2001}
Norris, J.~E., Ryan, S.~G., \& Beers, T.~C. 2001, ApJ, 561, 1034

\bibitem[{O'Brian {et~al.}(1991)O'Brian, Wickliffe, Lawler, Whaling,
  \& Brault}]{OBrianetal:1991} O'Brian, T. R., Wickliffe, M. E.,
  Lawler, J. E., Whaling, W., \& Brault, J. W. 1991,
  J.~Opt.~Soc.~Am.~B, 8, 1185

\bibitem[{Picardi {et~al.}(2004)Picardi, Chieffi, Limongi, Pisanti, Miele,
  Mangano, \& Imbriani}]{picardi04}
Picardi, I., Chieffi, A., Limongi, M., Pisanti, O., Miele, G., Mangano, G., \&
  Imbriani, G. 2004, ApJ, 609, 1035

\bibitem[{Pinsonneault {et~al.}(1999)Pinsonneault, Walker, Steigman, \&
  Narayanan}]{pinsonneault99}
Pinsonneault, M.~H., Walker, T.~P., Steigman, G., \& Narayanan, V.~K. 1999,
  \apj, 527, 180

\bibitem[{{Plez} \& {Cohen}(2005)}]{plez05}
{Plez}, B., \& {Cohen}, J.~G. 2005, \aap, 434, 1117

\bibitem[{Prantzos {et~al.}(1990)Prantzos, Hashimoto, \& Nomoto}]{prantzos90}
Prantzos, N., Hashimoto, M., \& Nomoto, K. 1990, A\&A, 234, 211

\bibitem[{Ram{\'{\i}}rez \& Mel{\' e}ndez(2004)}]{Ramirez/Melendez:2004}
Ram{\'{\i}}rez, I., \& Mel{\' e}ndez, J. 2004, ApJ, 609, 417

\bibitem[{Ryan {et~al.}(2002)Ryan, Gregory, Kolb, Beers, \&
  Kajino}]{Ryanetal:2002}
Ryan, S.~G., Gregory, S.~G., Kolb, U., Beers, T.~C., \& Kajino, T. 2002, ApJ,
  571, 501

\bibitem[{Ryan {et~al.}(1999)Ryan, Norris, \& Beers}]{Ryanetal:1999}
Ryan, S.~G., Norris, J.~E., \& Beers, T.~C. 1999, ApJ, 523, 654

\bibitem[{Schlegel {et~al.}(1998)Schlegel, Finkbeiner, \&
  Davis}]{Schlegeletal:1998} Schlegel, D.~J., Finkbeiner, D.~P., \&
  Davis, M. 1998, ApJ, 500, 525

\bibitem[{Shigeyama {et~al.}(2003)Shigeyama, Tsujimoto, \&
  Yoshii}]{Shigeyamaetal:2003}
Shigeyama, T., Tsujimoto, T., \& Yoshii, Y. 2003, ApJ, 586, L57

\bibitem[{{Shortridge}(1993)}]{figaro}
{Shortridge}, K. 1993, in ASP Conf. Ser. 52: Astronomical Data Analysis
  Software and Systems II, 219

\bibitem[{Sneden {et~al.}(2003)Sneden, Cowan, Lawler, Ivans, Burles, Beers,
  Primas, Hill, Truran, Fuller, Pfeiffer, \& Kratz}]{sneden03}
Sneden, C., et al. 2003, ApJ, 591, 936

\bibitem[{{Spergel} {et~al.}(2003){Spergel}, {Verde}, {Peiris},
  {Komatsu}, {Nolta}, {Bennett}, {Halpern}, {Hinshaw}, {Jarosik},
  {Kogut}, {Limon}, {Meyer}, {Page}, {Tucker}, {Weiland}, {Wollack},
  \& {Wright}}]{spergel03} {Spergel}, et al. 2003, \apjs, 148, 175

\bibitem[{{Spite} {et~al.}(2005){Spite}, {Cayrel}, {Plez}, {Hill},
  {Spite}, {Depagne}, {Fran{\c c}ois}, {Bonifacio}, {Barbuy}, {Beers},
  {Andersen}, {Molaro}, {Nordstr{\" o}m}, \& {Primas}}]{spite05}
  {Spite}, M., et al. 2005, \aap, 430, 655

\bibitem[{Stehl{\' e} \& {Hutcheon}(1999)}]{Stehle/Hutcheon:1999}
Stehl{\' e}, C., \& {Hutcheon}, R. 1999, A\&AS, 140, 93

\bibitem[{{Suda} {et~al.}(2004){Suda}, {Aikawa}, {Machida}, {Fujimoto}, \&
  {Iben}}]{suda04}
{Suda}, T., {Aikawa}, M., {Machida}, M.~N., {Fujimoto}, M.~Y., \& {Iben}, I.~J.
  2004, \apj, 611, 476

\bibitem[Surman \& McLaughlin (2005)]{surman05}
Surman, R., McLaughlin, G. C. 2005, ApJ, 618, 397

\bibitem[{Takeda {et~al.}(2003)Takeda, Zhao, Takada-Hidai, Chen, Saito, \&
  Zhang}]{takeda03}
Takeda, Y., Zhao, G., Takada-Hidai, M., Chen, Y.-Q., Saito, Y.-J., \& Zhang,
  H.-W. 2003, ChJAA, 3, 316

\bibitem[{Th\'{e}venin \& Idiart(1999)}]{thevenin99}
Th\'{e}venin, F., \& Idiart, T.~P. 1999, ApJ, 521, 753

\bibitem[{Travaglio {et~al.}(2004)Travaglio, Gallino, Arnone, Cowan, Jordan, \&
  Sneden}]{travaglio04}
Travaglio, C., Gallino, R., Arnone, E., Cowan, J., Jordan, F., \& Sneden, C.
  2004, ApJ, 601, 864

\bibitem[{Truran {et~al.}(2002)Truran, Cowan, Pilachowski, \&
  Sneden}]{truran02}
Truran, J.~W., Cowan, J.~J., Pilachowski, C.~A., \& Sneden, C. 2002, PASP, 114,
  1293

\bibitem[{Tsujimoto \& Shigeyama(2003)}]{Tsujimoto/Shigeyama:2003}
Tsujimoto, T., \& Shigeyama, T. 2003, ApJ, 584, L87

\bibitem[{Tsujimoto {et~al.}(1999)Tsujimoto, Shigeyama, \&
  Yoshii}]{Tsujimotoetal:1999}
Tsujimoto, T., Shigeyama, T., \& Yoshii, Y. 1999, ApJ, 519, L63

\bibitem[{Umeda \& Nomoto(2003)}]{Umeda/Nomoto:2003}
Umeda, H., \& Nomoto, K. 2003, Nature, 422, 871

\bibitem[{{Umeda} \& {Nomoto}(2005)}]{umeda05}
{Umeda}, H., \& {Nomoto}, K. 2005, \apj, 619, 427

\bibitem[{Vidal {et~al.}(1973)Vidal, Cooper, \& Smith}]{VCS}
Vidal, C. R., Cooper, J., \& Smith, E. W. 1973, ApJS, 25, 37

\bibitem[Wanajo \& Ishimaru (2005)]{wanajo05}Wanajo, S., \& Ishimaru,
Y. 2005, ASP Conf. Ser., IAU symposium 228, From Lithium to Uranium:
Elemental tracers of early cosmic evolution, in press

\bibitem[{Weiss {et~al.}(2004)Weiss, Schlattl, Salaris, \&
Cassisi}]{weiss04} Weiss, A. R., Schlattl, H., Salaris, M., \&
Cassisi, S. 2004, A\&A, 422, 217

\bibitem[{Wiese \& Martin(1980)}]{wiese80}
Wiese, W.~L., \& Martin, G.~A. 1980, NSRDS-NBS, 68

\bibitem[{Wisotzki {et~al.}(2000)Wisotzki, Christlieb, Bade, Beckmann,
  K\"ohler, Vanelle, \& Reimers}]{hespaperIII}
Wisotzki, L., Christlieb, N., Bade, N., Beckmann, V., K\"ohler, T., Vanelle,
  C., \& Reimers, D. 2000, A\&A, 358, 77

\bibitem[Yoshii (1981)]{yoshii81} Yoshii, Y. 1981, \aap, 97, 280

\bibitem[{Yoshii {et~al.}(2003)Yoshii, Kobayashi, \&
  Minezaki}]{Yoshiietal:2003}
Yoshii, Y., Kobayashi, Y., \& Minezaki, T. 2003, BAAS, 202, 38.03

\bibitem[{Zachwieja(1995)}]{zachwieja95}
Zachwieja, M. 1995, J. Mol. Spectrosc., 170, 285

\bibitem[{Zachwieja(1997)}]{zachwieja97}
---. 1997, J. Mol. Spectrosc., 182, 18

\end{thebibliography}




\clearpage
\begin{deluxetable}{lllrll} 
\tablecolumns{7} 
\tablewidth{0pt} 
\tablecaption{\label{Tab:SubaruObs} Subaru Observations of {\FAC} and {\GG}.} 
\tablehead{
  \colhead{Target} & \colhead{UT\tablenotemark{a}} & \colhead{Setting} & \colhead{$t$} & \colhead{$v_{\rm r}$} & \colhead{Notes}\\
  \colhead{} & \colhead{} & \colhead{} & \colhead{(min)} & \colhead{km~s$^{-1}$} & \colhead{}
  }
\startdata 
 HE~1327$-$2326 & 2004 May 30,  7:04  & 4030--6800\,{\AA} &  30 &        & Close to the moon\\
 HE~1327$-$2326 & 2004 May 31,  6:17  & 4030--6800\,{\AA} & 150 & 63.88  & Used for abundance analysis \\
 HE~1327$-$2326 & 2004 June 2,  5:43  & 3550--5250\,{\AA} & 240 & 63.50  & Used for abundance analysis \\
 HE~1327$-$2326 & 2004 June 27, 6:03  & 3000--4600\,{\AA} & 150 & 63.64  & Used for abundance analysis \\
 HE~1327$-$2326 & 2005 Feb. 27, 12:48 & 4030--6800\,{\AA} &  30 & 63.62  & \\
 HE~1327$-$2326 & 2005 June 17, 5:46  & 4030--6800\,{\AA} &  15 & 63.45  & \\
 {\GG}         & 2004 June 2,  9:52  & 3550--5250\,{\AA} &  20 & 443.84 & $=$ HE~1337$+$0012\\
 {\GG}         & 2004 June 27, 8:42  & 3000--4600\,{\AA} &  60 & 443.67 & $=$ HE~1337$+$0012\\
\enddata
\tablenotetext{a}{At beginning of observation}
\end{deluxetable}

\begin{deluxetable}{lllrrrrrl} 
\tablecolumns{9} 
\tablewidth{0pt} 
\tablecaption{\label{Tab:Eqw} Atomic Data and Measured Equivalent Widths}
\tablehead{
  \colhead{} & \colhead{} & \colhead{} & \colhead{} & \multicolumn{4}{c}{$W_{\lambda}$}\\
  \cline{5-8}\\
  \colhead{}    & \colhead{$\lambda$} & \colhead{$\chi$} & \colhead{$\log gf$} &
  \colhead{W.A.} & \colhead{A.F.} & \colhead{$\Delta$} & \colhead{Adopted} & \colhead{}\\
  \colhead{Ion} & \colhead{({\AA})}    & \colhead{(eV)}  & \colhead{(dex)}  &
  \colhead{(m{\AA})} & \colhead{(m{\AA})} & \colhead{(m{\AA})} & \colhead{(m{\AA})} & 
  \colhead{Refs.}
  }
\startdata
\ion{Na}{1} & $5889.951$ & $0.000$ & $ 0.117$ & $ 48.9$ & \nodata & \nodata & $ 48.9$ & 1 \\
\ion{Na}{1} & $5895.924$ & $0.000$ & $-0.184$ & $ 31.5$ & \nodata & \nodata & $ 31.5$ & 1 \\
\ion{Mg}{1} & $3829.355$ & $2.707$ & $-0.208$ & $ 22.6$ & $ 22.9$ & $-0.3$  & $ 22.8$ & 1 \\
\ion{Mg}{1} & $5167.321$ & $2.709$ & $-1.030$ & $  9.8$ & \nodata & \nodata & $  9.8$ & 1 \\
\ion{Mg}{1} & $5172.684$ & $2.712$ & $-0.402$ & $ 21.1$ & $ 23.5$ & $-2.4$  & $ 22.3$ & 1 \\
\ion{Mg}{1} & $5183.604$ & $2.717$ & $-0.180$ & $ 30.1$ & $ 30.8$ & $-0.7$  & $ 30.5$ & 1 \\
\ion{Al}{1} & $3961.529$ & $0.014$ & $-0.336$ & $ 11.0$ & $ 13.0$ & $-2.0$  & $ 12.0$ & 2 \\
\ion{Ca}{1} & $4226.728$ & $0.000$ & $ 0.244$ & $  2.7$ & \nodata & \nodata & $  2.7$ & 2 \\
\ion{Ca}{2} & $3933.663$ & $0.000$ & $ 0.105$ & $128.9$ & $135.5$ & \nodata & $132.2$ & 1 \\
\ion{Ti}{2} & $3234.520$ & $0.049$ & $ 0.426$ & $  7.2$ & \nodata & \nodata & $  7.2$ & 1 \\   
\ion{Ti}{2} & $3349.408$ & $0.049$ & $ 0.586$ & $ 16.6$ & \nodata & \nodata & $ 16.6$ & 1 \\
\ion{Ti}{2} & $3759.300$ & $0.607$ & $ 0.270$ & $  5.9$ & \nodata & \nodata & $  5.9$ & 3 \\
\ion{Ti}{2} & $3761.330$ & $0.574$ & $ 0.170$ & $  4.7$ & \nodata & \nodata & $  4.7$ & 3 \\
\ion{Fe}{1} & $3581.193$ & $0.859$ & $ 0.415$ & $  5.9$ & $  6.8$ & $-0.9$  & $  6.4$ & 4 \\
\ion{Fe}{1} & $3737.131$ & $0.052$ & $-0.572$ & $  3.9$ & \nodata & \nodata & $  3.9$ & 4 \\
\ion{Fe}{1} & $3745.561$ & $0.087$ & $-0.767$ & $  4.8$ & \nodata & \nodata & $  4.8$ & 4 \\
\ion{Fe}{1} & $3758.233$ & $0.958$ & $-0.005$ & $  5.1$ & \nodata & \nodata & $  5.1$ & 4 \\
\ion{Fe}{1} & $3820.425$ & $0.859$ & $ 0.158$ & $  2.5$ & \nodata & \nodata & $  2.5$ & 4 \\
\ion{Fe}{1} & $3859.912$ & $0.000$ & $-0.710$ & $  6.8$ & $  5.9$ & $+0.9$  & $  6.4$ & 5\\
\ion{Fe}{1} & $4045.812$ & $1.485$ & $ 0.285$ & $  1.9$ & \nodata & \nodata & $  1.9$ & 4 \\
\ion{Sr}{2} & $4077.724$ & $0.000$ & $ 0.158$ & $  7.3$ & $  5.0$ & $+2.3$  & $  6.2$ & 6 \\
\ion{Sr}{2} & $4215.540$ & $0.000$ & $-0.155$ & $  3.8$ & \nodata & \nodata & $  3.8$ & 6 \\
\ion{Cr}{1} & $4254.332$ & $0.000$ & $-0.114$ & $ <2.0$ & \nodata & \nodata & $ <2.0$ & 3  \\
\ion{Mn}{1} & $4030.753$ & $0.000$ & $-0.470$ & $ <2.0$ & \nodata & \nodata & $ <2.0$ & 3  \\
\ion{Fe}{2} & $5018.450$ & $2.891$ & $-1.220$ & $ <2.0$ & \nodata & \nodata & $ <2.0$ & 4 \\
\ion{Co}{1} & $3453.514$ & $0.432$ & $ 0.380$ & $ <6.3$ & \nodata & \nodata & $ <6.3$ & 4 \\
\ion{Ni}{1} & $3414.761$ & $0.025$ & $-0.029$ & $ <7.5$ & \nodata & \nodata & $ <7.5$ & 7 \\
\ion{Zn}{1} & $4810.530$ & $4.080$ & $-0.150$ & $ <2.0$ & \nodata & \nodata & $ <2.0$ & 8 \\
\ion{Ba}{2} & $4554.029$ & $0.000$ & $ 0.170$ & $ <1.8$ & \nodata & \nodata & $ <1.8$ & 1\\
\enddata
\tablerefs{(1) VALD \citep{vald99}; (2) \citet{wiese80}; (3) \citet{morton91};
(4)\citet{OBrianetal:1991}; (5)\citet{Fuhretal:1988}; (6)\citet{grevesse81};
 (7)\citet{blackwell89}; (8)\citet{kerkhoff80}}

\end{deluxetable} 

\begin{deluxetable}{lcccccccc} 
\tablewidth{0pt} 
\tablecaption{\label{tab:lw} Line widths measurements}
\tablehead{
  \colhead{} & \multicolumn{4}{c}{\FAC} & & \multicolumn{3}{c}{\GG} \\
  \cline{2-5} \cline{7-9} 
  \colhead{} & W$_{\lambda}$ & FWHM & \multicolumn{2}{c}{$v_{\rm macro+inst}$ (km~s$^{-1}$)} & & W$_{\lambda}$ & FWHM & $v_{\rm macro+inst}$ \\
  \cline{4-5}
  \colhead{} & m{\AA} & km~s$^{-1}$ & dwarf & subgiant & & m{\AA} & km~s$^{-1}$ & km~s$^{-1}$ 
}
\startdata
  \ion{Mg}{1} 5172~{\AA}       & 22.3 & 9.45 & 6.22 & 6.31 & & 78.6 & 9.62 & 4.65  \\
  \ion{Mg}{1} 5183~{\AA}       & 30.5 & 8.91 & 5.43 & 5.52 & & 92.8 & 10.07 & 4.56  \\
\enddata
\end{deluxetable} 

\begin{deluxetable}{cclllllrrlc} 
\rotate
\tabletypesize{\footnotesize}
\tablecolumns{11} 
\tablewidth{0pt} 
\tablecaption{\label{Tab:ISCa}Results of VPFIT for interstellar \ion{Ca}{2} K Lines of {\FAC}} 
\tablehead{
  \colhead{}&  \colhead{Component} & \colhead{$\log N$} & \colhead{$\sigma_{\log N}$} &
  \colhead{$v_{\rm helio}$} &  \colhead{$\sigma_{v}$} & 
  \colhead{$b$} &  \colhead{$\sigma_{b}$} & 
  \colhead{W\tablenotemark{a}}   & \colhead{Remark} &
  \colhead{Corresponding}\\
  \multicolumn{2}{l}{} & \colhead{(cm$^{-2}$)} & 
  \multicolumn{1}{l}{} & \colhead{(km/s)} & 
  \multicolumn{1}{l}{} & \colhead{(km/s)} & 
  \multicolumn{1}{l}{} & \colhead{(m\,\AA)} & 
  \multicolumn{1}{l}{} & \colhead{\ion{Na}{1} D component}
}
\startdata 
 &  1 &  11.6593  &  0.0620 &  $+$33.2 & 0.6   & 3.40  & 0.77   & 38.8  & Gaussian fitting	     &   1 \\
 &  2 &  11.4444  &  0.1431 &  $+$26.3 & 0.9   & 3.48  & 0.60   & 19.1  & Direct integral &   2 \\
 &  3 &  11.5587  &  0.0540 &  $+$17.6 & 0.3   & 3.44  & 0.55   & 32.6  & Gaussian fitting	    &    3 \\
 &  4 &  10.9804  &  0.5561 &  $+$8.9 & 2.7   & 4.72  & 5.63   & 17.4 & Direct integral for component 4 and 5 &   4 \\
 &  5 &  11.3510  &  0.4160 &  $-$1.3 & 1.8   & 6.06  & 4.74   & \nodata   & \nodata  &   5 \\
 &  6 &  11.8168  &  0.0917 &  $-$11.2 & 0.9   & 6.21  & 0.28   & 64.2  & Gaussian fitting	     &   5 \\
 &  7 &  11.0830  &  0.0258 &  $-$29.8 & 0.3   & 3.92  & 0.37   & 11.6  & Gaussian fitting             &   6+7 \\
\enddata
\tablenotetext{a}{A total equivalent width of 179.5\,m\,{\AA} is measured from a direct
integration between 3932.27 and 3933.40\,{\AA}.}
\end{deluxetable}



\clearpage
\thispagestyle{empty}
\begin{deluxetable}{cclllllrrrlc} 
\rotate
\tabletypesize{\footnotesize}
\tablecolumns{10} 
\tablewidth{0pt} 
\tablecaption{\label{Tab:ISNa}Results of
VPFIT for interstellar \ion{Na}{1} D Lines of {\FAC}} 
\tablehead{ 
\colhead{}&
\colhead{Component} & \colhead{$\log N$} & \colhead{$\sigma_{\log N}$} &
\colhead{$v_{\rm helio}$\tablenotemark{a}} & \colhead{$\sigma_{v}$} & 
\colhead{$b$} & \colhead{$\sigma_{b}$} & 
\colhead{W$_{\rm D2}$\tablenotemark{b}}   & \colhead{W$_{\rm D1}$}   & \colhead{Remark} &
\colhead{Corresponding}\\ 
\multicolumn{2}{l}{} & \colhead{(cm$^{-2}$)} & 
\multicolumn{1}{l}{} & \colhead{(km/s)} &
\multicolumn{1}{l}{} & \colhead{(km/s)} & 
\multicolumn{2}{l}{} & \colhead{(m\,\AA)} & 
\multicolumn{1}{l}{} & \colhead{\ion{Ca}{2} K component} 
} 
\startdata 

& 1 & 11.8212 & 0.1493 & $+$32.6 & 0.6 & 2.80 & 0.93 & 108.3 & 79.9 & Direct integral for compoment 1 and 2   & 1 \\ 

& 2 & 11.3985 & 0.6202 & $+$26.6 & 3.9 & 4.76 & 5.21 & \nodata & \nodata &  & 2 \\ 
& 3 & 11.5795 & 0.1426 & $+$17.6 & 0.9 & 2.85 & 0.56 & 61.4 & 39.2 & Gaussian fitting	& 3 \\ 
& 4 & 10.8796 & 0.1873 & $+$8.6 & 1.8 & 3.79 & 2.00 &  14.5 & 13.1 & Gaussian fitting  & 4 \\ 
& 5 & 11.9590 & 0.0140 & $-$7.0 & 0.0 & 4.72 & 0.20 & 128.3 & 77.1 & Direct integral & 5+6 \\
& 6 & 10.4562 & 1.2197 & $-$27.4 & 8.1 & 2.67 &13.56 &  5.7 & 3.6 & Gaussian fitting for D2~\tablenotemark{c}    & 7 \\ 
& 7 & 10.8254 & 0.5790 & $-$34.6 & 5.4 & 3.91 & 2.77 & 14.6 & \nodata & Gaussian fitting for D2~\tablenotemark{c}  & 7\\
\enddata

\tablenotetext{a}{The velocity was measured with a simultaneous fitting for both
  lines.}


\tablenotetext{b}{A total equivalent width of 312.8\,m\,{\AA} is measured from a
direct integration between 5888.35 and 5889.54\,\AA (features 1--5) and
between 5887.88 and 5888.27\,\AA (features 6--7).}

\tablenotetext{c}{A direct integral was applied to the component 6 and 7 of the D1 line.} 

\end{deluxetable}

\begin{deluxetable}{lrrrrrrrrr} 
\tablecolumns{10} 
\tablewidth{0pt}
\tablecaption{\label{Tab:Photometry} Broadband Optical and Infrared Photometry of
  {\FAC} and {\GG}} 
\tablehead{
  \colhead{} & \colhead{$U$} & \colhead{$B$} & \colhead{$V$} & \colhead{$R$} &
  \colhead{$I$} & \colhead{$J$} & \colhead{$H$} & \colhead{$K_S$} \\
  \colhead{Star} & \colhead{(mag)} & \colhead{(mag)} & \colhead{(mag)} &
  \colhead{(mag)} & \colhead{(mag)} & \colhead{(mag)} & \colhead{(mag)} & \colhead{(mag)}
  }
\startdata
{\FAC}\tablenotemark{a} &  13.787 & 14.016 & 13.535 & 13.211 & 12.854 & 12.357 & 12.068 & 11.986\\
$\sigma$                &   0.007 &  0.005 &  0.004 &  0.003 &  0.003 &  0.023 &  0.023 &  0.021\\
{\FAC}\tablenotemark{b} & \nodata & 14.01  & 13.53  & 13.20  & \nodata & \nodata & \nodata & \nodata \\
$\sigma$                & \nodata &  0.04  &  0.02  &  0.06  & \nodata & \nodata & \nodata & \nodata \\
{\FAC}\tablenotemark{c} &  13.761 & 13.966 & 13.526 & 13.156 & 12.803 & \nodata & \nodata & \nodata \\
$\sigma$                &   0.019 &  0.008 &  0.018 &  0.013 &  0.021 & \nodata & \nodata & \nodata \\
{\GG}\tablenotemark{d} &  11.601 & 11.838 & 11.453 & 11.186 & 10.893 & 10.509 & 10.268 & 10.208\\
$\sigma$                &  0.0044 & 0.0034 & 0.0016 & 0.0021 & 0.0022 &  0.024 &  0.023 &  0.021\\
{\GG}\tablenotemark{c} &  11.599 & 11.769 & 11.442 & 11.156 & 10.859 & \nodata & \nodata & \nodata \\
$\sigma$                &  0.021 & 0.015 & 0.009 & 0.006 & 0.007 & \nodata & \nodata & \nodata \\
\enddata

\tablenotetext{a}{$UBVRI$ measurements obtained differentially to
  {\GG} with the MAGNUM telescope. $JHK$ measurements are from
  2MASS. These values were used to estimate the effective
  temperature.}
\tablenotetext{b}{$BVR$ measurements with the WIYN telescope.}
\tablenotetext{c}{$UBVRI$ measurements with the CTIO telescope.}
\tablenotetext{d}{$UBVRI$ is from \cite{Landolt:1992}; $JHK$ from 2MASS.}
\end{deluxetable} 

\begin{deluxetable}{lllccccl} 
\tablecolumns{9} 
\tabletypesize{\footnotesize}
\tablewidth{0pt} 
\tablecaption{\label{Tab:TeffDerivation} Derivation of {\tefft} for
  HE~1327$-$2326 and {\GG}.} 
\tablehead{
  \colhead{} & \multicolumn{2}{c}{Value} & \colhead{} & \multicolumn{2}{c}{Derived {\tefft}} & 
  \colhead{} & \colhead{} \\
  \cline{2-3} \cline{5-6} \\ 
  \colhead{Measured} & \colhead{\FAC} & \colhead{\GG} & \colhead{} & \colhead{\FAC} & \colhead{\GG} & 
  \colhead{$\Delta\teffm$} & \colhead{} \\
  \colhead{quantity} & \colhead{}  & \colhead{} & \colhead{}  & \colhead{(K)} & 
  \colhead{(K)} & \colhead{(K)} & \colhead{Notes}           
  }
\startdata
 H$\alpha$--H$\delta$ & \nodata & \nodata  & & $5990$ & $6300$ & $310$ & STEHLE$+$BPO; $\log g=4.5, 4.4$~\tablenotemark{a} \\ 
 H$\alpha$--H$\delta$ & \nodata & \nodata  & & $6050$ & $6300$ & $250$ & STEHLE$+$BPO; $\log g=3.7, 4.4$~\tablenotemark{a} \\ 
 H$\alpha$--H$\delta$ & \nodata & \nodata  & & $6040$ & $    $ & $   $ & VCS$+$AG; $\log g=3.7$~\tablenotemark{b} \\[1ex]
\hline
 HP2                  & $4.01$~{\AA}  & $4.77$~{\AA}   & & $6000$ & $6200$ & $200$ & \citet{Ryanetal:1999} \\
 HP2                  & $4.01$~{\AA}  & $4.77$~{\AA}   & & $6160$ & $6350$ & $190$ & Beers et al. (2005, in preparation)  \\
\hline
 $(B-V)_{0}$     & 0.403\,mag & 0.357\,mag & & $6130$ & $6340$ & $210$ & \citet{Alonsoetal:1996}~\tablenotemark{c} \\
 $(V-R)_{0}$     & 0.413\,mag & 0.376\,mag & & $6290$ & $6520$ & $230$ & \citet{Alonsoetal:1996}~\tablenotemark{c} \\
 $(V-I)_{0}$     & 0.726\,mag & 0.658\,mag & & $6170$ & $6430$ & $260$ & \citet{Alonsoetal:1996}~\tablenotemark{c} \\
 $(V-K)_{0}$     & 1.323\,mag & 1.166\,mag & & $6130$ & $6440$ & $310$ & \citet{Alonsoetal:1996}~\tablenotemark{c} \\
 Average         &            &            & & $6180$ & $6430$ & $250$ & 
\enddata

\tablenotetext{a}{Computations of Stark broadening by
    \citet{Stehle/Hutcheon:1999} and self-broadening by
    \citet{Barklemetal:2000} were employed for the analysis. The $\log
    g$ values for {\FAC} and {\GG} adopted in this analysis are
    given.}

\tablenotetext{b}{Korn (2004; private communication). Computations of
    Stark broadening by \citet{VCS} and self-broadening by
    \citet{Ali/Griem:1966} were employed for the analysis. The $\log
    g$ value for {\FAC} adopted in this analysis is given.}

\tablenotetext{b}{Their calibration scale for $\mbox{[Fe/H]}=-3.0$ was
used.}

\end{deluxetable}

\begin{deluxetable}{llll} 
\tabletypesize{\footnotesize}
\tablecolumns{4} 
\tablewidth{0pt} 
\tablecaption{\label{Tab:StellarParameters} Stellar Parameters.} 
\tablehead{
  \colhead{Parameter} & \multicolumn{2}{c}{\FAC} & {\GG} \\
\cline{2-3}
  \colhead{} & \colhead{Subgiant} & \colhead{Dwarf} & 
  }
\startdata 
 $T_{\mbox{\scriptsize eff}}$  & $6180$\,K   & $6180$\,K  & 6390~K \\
 $\log g$ (cgs)                & $3.7$  & $4.5$ & 4.38 \\
 $\mbox{[Fe/H]}$               & $-5.6$ & $-5.7$  & $-3.2$ \\
 $v_{\mbox{\scriptsize micr}}$ & $1.7$\,km\,s$^{-1}$ & $1.5$\,km\,s$^{-1}$ & $1.6$\,km\,s$^{-1}$ \\
\enddata
\end{deluxetable} 

\clearpage
\thispagestyle{empty}
\begin{deluxetable}{lcccrrrrcrrrrccrrrl} 
\rotate
\tabletypesize{\tiny}
\tablewidth{0pc} 
\tablecaption{\label{Tab:Abundances} Abundances of {\FAC} and {\GG}} 
\tablehead{ 
\colhead{} & \colhead{} & \colhead{} & 
\multicolumn{10}{c}{\FAC} & & 
\multicolumn{4}{c}{\GG} &  \colhead{} \\
\cline{4-13} \cline{15-19} 
\colhead{} & \colhead{} & \colhead{} & \colhead{} & &
\multicolumn{3}{c}{Subgiant} & \colhead{} & \multicolumn{3}{c}{Dwarf} & & &
\colhead{} & \colhead{} & \colhead{} &  \colhead{} \\ 
\cline{5-8} \cline{10-13} 
\colhead{El.} & \colhead{Ion} &   
\colhead{$\log\epsilon (\mbox{X})_{\odot}$} & \colhead{$N_{\mbox{\scriptsize lines}}$} &
\colhead{$\log\epsilon (\mbox{X})$} & \colhead{[X/H]} & \colhead{[X/Fe]} & \colhead{$\log\epsilon (\mbox{X})_{\rm F}$}\tablenotemark{a} & & 
\colhead{$\log\epsilon (\mbox{X})$} & \colhead{[X/H]} & \colhead{[X/Fe]} & \colhead{$\log\epsilon (\mbox{X})_{\rm F}$}\tablenotemark{a} & & 
\colhead{$N_{\mbox{\scriptsize lines}}$} &
\colhead{$\log\epsilon (\mbox{X})$} & \colhead{[X/H]} & \colhead{[X/Fe]} & 
\colhead{Notes}}
\startdata 
C  & CH & $8.39$ & 3  & $ 6.99$ & $ -1.40$ & $ 4.26$ & 6.90 && $ 6.79$ & $ -1.60$ & $ 3.90$ & 6.64   && & 5.68\tablenotemark{b} & $-2.71$ & $+0.49$ & Synth. of CH    \\
N  & NH & $7.93$ & 1  & $ 6.83$ & $ -1.10$ & $ 4.56$ & 6.68 && $ 6.33$ & $ -1.60$ & $ 4.08$ & 6.36   && & 6.15\tablenotemark{c} & $-1.78$ & $+1.42$ & Synth. of NH    \\
O  & OH & $8.66$ & 1  &   & $<-1.66$ & $<4.0$ &$<7.0$&& \nodata & $<-1.96$ & $<3.69$ &$<-1.96$&&   & 6.34\tablenotemark{b} & $-2.32$ & $+0.88$ & Synth. of OH  \\
Li & 1  & $1.16$ & 1  & $<1.5 $ & \nodata  & \nodata & $<1.6$ && $<1.5 $ & \nodata  & \nodata &$<1.6$  && 1 & 2.30\tablenotemark{d} & \nodata & \nodata & Synthesis   \\
Na & 1  & $6.33$ & 2  & $ 3.06$ & $ -3.11$ & $ 2.55$ & 2.92&& $ 3.06$ & $ -3.11$ & $ 2.54$ & 2.95 && 1 & 2.10\tablenotemark{d} & $-4.23$ & $-1.13$ & \\
   &    &        &    & $ 2.86$ & $ -3.31$ & $ 2.15$ & 2.72&& $ 2.86$ & $ -3.31$ & $ 2.14$ & 2.75 &&   & 1.90                  & $-4.43$ & $-1.53$ & non-LTE: $-0.2$ \\
Mg & 1  & $7.54$ & 4  & $ 3.63$ & $ -3.90$ & $ 1.76$ & 3.57&& $ 3.64$ & $ -3.89$ & $ 1.76$ & 3.57 && 5 & $ 4.80$ & $-2.74$ & $+0.46$ &  \\
   &    &        &    & $ 3.73$ & $ -3.80$ & $ 1.66$ & 3.67&& $ 3.74$ & $ -3.79$ & $ 1.66$ & 3.67 &&   & $ 4.90$ & $-2.64$ & $+0.36$ & non-LTE: $+0.1$ \\
Al & 1  & $6.47$ & 1  & $ 2.04$ & $ -4.33$ & $ 1.33$ & 2.05&& $ 2.05$ & $ -4.32$ & $ 1.33$ & 2.01 && 1 & $ 2.75$ & $-3.72$ & $-0.52$ &    \\
   &    &        &    & $ 2.64$ & $ -3.73$ & $ 1.73$ & 2.65&& $ 2.65$ & $ -3.72$ & $ 1.73$ & 2.61 &&   & $ 3.25$ & $-3.12$ & $-0.12$ & non-LTE: $+0.6$\\
Ca & 2  & $6.36$ & 1  & $ 1.52$ & $ -4.79$ & $ 0.87$ & 1.57&& $ 1.50$ & $ -4.81$ & $ 0.84$ & 1.45 && \nodata & \nodata &  \nodata & \nodata &   \\
Ca & 1  & $6.36$ & 1  & $ 0.95$ & $ -5.36$ & $ 0.30$ & 0.95&& $ 0.97$ & $ -5.34$ & $ 0.31$ & 0.94 && 6 & $ 3.62$ & $-2.74$ & $+0.46$ &  \\
Ti & 2  & $5.02$ & 2  & $-0.24$ & $ -5.14$ & $ 0.52$ &$-0.17$ && $ 0.02$ & $ -4.88$ & $ 0.77$ & 0.03 && 58 & $ 2.20$ & $-2.82$ & $+0.38$ &  \\
Fe & 1  & $7.45$ & 4  & $ 1.79$ & $ -5.66$ & \nodata & 1.83&& $ 1.80$ & $ -5.65$ & \nodata & 1.80 && 59 & $ 4.25$ & $-3.20$ & \nodata &  \\
Fe & 1  & $7.45$ & 4  & $ 1.99$ & $ -5.46$ & \nodata & 2.03&& $ 2.00$ & $ -5.45$ & \nodata & 2.00 &&    & $ 4.45$ & $-3.00$ & \nodata & non-LTE: $+0.2$\\
Sr & 2  & $2.92$ & 2  & $-1.77$ & $ -4.69$ & $ 0.97$ &$-1.75$ && $-1.49$ & $ -4.41$ & $ 1.24$ &$-1.50$&& 2 & $-0.10$ & $-3.02$  & $+0.18$ &  \\
Sr & 2  & $2.92$ & 2  & $-1.47$ & $ -4.39$ & $ 1.07$ &$-1.45$ && $-1.19$ & $ -4.11$ & $ 1.34$ &$-1.20$&& 2 & $ 0.20$ & $-2.72$  & $+0.28$ & non-LTE: $+0.3$\\
Cr & 1  & $5.64$ &\nodata& $<1.09$ & $<-4.55$ & $<1.11$ &  \nodata & &  $< 1.11$ & $<-4.53$ & $<1.12$ &\nodata && 3 & $ 2.26$ & $-3.38$  & $-0.18$ & \\
Mn & 1  & $5.39$ &\nodata& $<0.87$ & $<-4.52$ & $<1.14$ &  \nodata & &  $< 0.89$ & $<-4.50$ & $<1.15$ &\nodata && 3 & $ 1.49$ & $-3.90$  & $-0.70$ & \\
Fe & 2  & $7.45$ &\nodata& $<3.01$ & $<-4.44$ & $<1.22$ &  \nodata & &  $< 3.31$ & $<-4.14$ & $<1.51$ &\nodata && 3 & $ 4.38$ & $-3.07$  & $+0.13$ & \\
Co & 1  & $4.92$ &\nodata& $<1.34$ & $<-3.58$ & $<2.08$ &  \nodata & &  $< 1.33$ & $<-3.59$ & $<2.06$ &\nodata && 12 & $ 2.19$ & $-2.73$  & $+0.47$ & \\
Ni & 1  & $6.23$ &\nodata& $<1.16$ & $<-5.07$ & $<0.59$ &  \nodata & &  $< 1.15$ & $<-5.08$ & $<0.57$ &\nodata && 24 & $ 3.04$ & $-3.19$  & $+0.01$ & \\
Zn & 1  & $4.60$ &\nodata& $<2.02$ & $<-2.49$ & $<3.07$ &  \nodata & &  $< 2.11$ & $<-2.49$ & $<3.16$ &\nodata && \nodata& \nodata & \nodata  & \nodata & \\
Ba & 2  & $2.13$ & 1  &$<-2.06$ & $<-4.20$ & $<1.46$ &$<-2.03$ &&$<-1.78$ & $<-3.95$ & $<1.70$ &$<-1.78$ && 2 & $-1.32$ & $-3.45$  & $-0.25$ & \\
Ba & 2  & $2.13$ & 1  &$<-1.86$ & $<-4.00$ & $<1.46$ &$<-1.83$ &&$<-1.58$ & $<-3.75$ & $<1.70$ &$<-1.58$ && 2 & $-1.12$ & $-3.25$  & $-0.25$ & non-LTE: $+0.2$\\
\enddata 
\tablenotetext{a}{Frebel et al. (2005)}
\tablenotetext{b}{Near infrared measurements by \citet{akerman04};}
\tablenotetext{c}{\citet{israelian04};}
\tablenotetext{d}{Aoki et al. (2005, in preparation)}
\end{deluxetable}

\begin{deluxetable}{lccccc} 
\tablewidth{0pc} 
\tablecaption{\label{tab:error} ERROR ESTIMATES} 
\tablehead{ 
Species & random & $\Delta T_{\rm eff}$ & $\Delta \log g$ & $\Delta v_{\rm turb}$ & r.s.s. \\
        &        &  +100~K               & +0.3~dex        & +0.3~km~s$^{-1}$  & 
}
\startdata 
C (CH)      & 0.1  & 0.20 & -0.10 & 0.00 & 0.24 \\
N (NH)      & 0.2  & 0.20 & -0.10 & 0.00 & 0.30 \\
\ion{Na}{1} & 0.04 & 0.07 & -0.01 &-0.02 & 0.08 \\    
\ion{Mg}{1} & 0.02 & 0.06 &  0.00 &-0.01 & 0.07 \\ 
\ion{Al}{1} & 0.03 & 0.08 &  0.00 &-0.01 & 0.09 \\ 
\ion{Ca}{1} & 0.15 & 0.08 &  0.00 & 0.00 & 0.17 \\ 
\ion{Ti}{2} & 0.17 & 0.07 &  0.08 &-0.01 & 0.20 \\ 
\ion{Fe}{1} & 0.11 & 0.10 &  0.00 & 0.00 & 0.15 \\ 
\ion{Sr}{2} & 0.06 & 0.06 &  0.09 & 0.00 & 0.13 \\ 
\enddata 
\end{deluxetable} 
\clearpage

\begin{figure}[htbp]
  \includegraphics[width=9cm]{f12.ps}
  \caption{\label{fig:xh} Chemical abundance patterns of {\FAC}
  (filled circles) and {\CBB} (open circles). The line with open
  circles indicates the abundance pattern of the average of extremely
  metal-poor stars with [Fe/H]$<-3.5$ ({\cd}, CS~22885--096,
  BS~16467--062, and CS~22172--002: \citet{Francoisetal:2003};
  \citet{Cayreletal:2004}), while the line with filled circles means
  that of the two carbon-rich objects (CS~22949--037 and
  CS~29498--043, Depagne et al. 2002; Aoki et al. 2004).}
\end{figure}

\begin{figure}[htbp]
  \includegraphics[width=9cm]{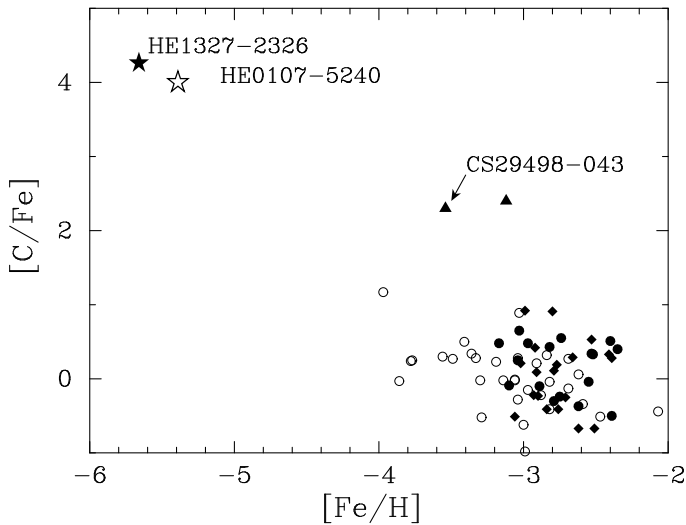}
  \caption{\label{fig:cfe} Carbon abundance ratio as a function of
  [Fe/H]. The filled star indicates the values of {\FAC} determined by
  the present work. Results from previous studies are shown by an open
  star \citep{HE0107_ApJ}, filled trianges \citep{Aokietal:2002d}, open
  circles \citep{Cayreletal:2004}, filled stars
  \citep{Hondaetal:2004b}, and filled diamonds \citep{aoki05}.}
\end{figure}

\begin{figure}[htbp]
  \includegraphics[width=9cm]{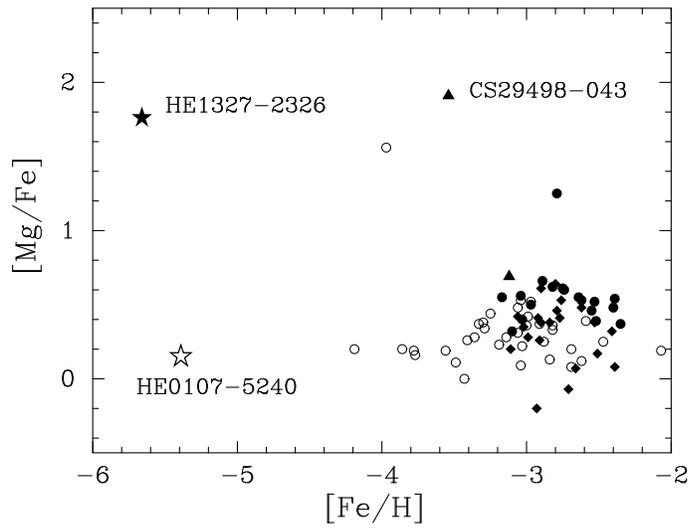}
  \caption{\label{fig:mgfe} Same as Fig.~\ref{fig:cfe}, but for Mg
  abundance ratio.}
\end{figure}
\end{document}